\documentclass[galaxies,article,accept,moreauthors,pdftex,12pt,a4paper]{mdpi}
\usepackage{amsmath}
\usepackage{amsfonts}
\usepackage{amssymb}
\usepackage{graphicx}
\usepackage[titletoc]{appendix}
\usepackage{color,soul,booktabs,subfigure}
\graphicspath{{figures/}}

\makeatletter
\renewcommand{\@thesubfigure}{\normalsize(\textbf{\alph{subfigure}})}
\makeatother

\newcommand{\abs}[1]{\vert{#1}\vert}

\graphicspath{{figures/}}
\usepackage{enumitem}
\setitemize{parsep=0pt,itemsep=0pt,leftmargin=.4in}
\setenumerate{parsep=0pt,itemsep=0pt,leftmargin=.4in}

\lastpage{260}
\doinum{10.3390/galaxies1030216}
\pubvolume{1}
\pubyear{2013}
\history{Received: 10 September 2013; in revised form: 6 November 2013 / Accepted: 12 November 2013 / \\
Published: 3 December 2013}

\Title{Cosmographic Constraints and Cosmic Fluids}

\Author{Salvatore Capozziello $^{1,2,}$*, Mariafelicia De Laurentis $^{1,2}$, Orlando Luongo $^{1,2,3}$ and \mbox{Alan Cosimo Ruggeri $^{1,2}$}}

\address{%
$^{1}$ Dipartimento di Fisica, Universit\`a di Napoli ``Federico II'', Via Cinthia, Napoli I-80126, Italy; \linebreak E-Mails: felicia@na.infn.it (M.L.); luongo@na.infn.it (O.L.); acruggeri@na.infn.it (A.C.R.)\\
$^{2}$ Istituto Nazionale di Fisica Nucleare (INFN), Sez. di Napoli, Via Cinthia, Napoli I-80126, Italy\\
$^{3}$ Instituto de Ciencias Nucleares, Universidad Nacional Autonoma de M\'exico (UNAM), \linebreak AP 70543, M\'exico, DF 04510, Mexico}

\corres{E-Mail: salvatore.capozziello@na.infn.it; \linebreak Tel.: +39-081-676-496; Fax: +39-081-676-352.}

\abstract{The problem of reproducing dark energy effects is reviewed here with particular interest devoted to cosmography. We summarize some of the most relevant cosmological models, based on the assumption that the corresponding barotropic equations of state evolve as the universe expands, giving rise to the accelerated expansion. We describe in detail the $\Lambda$CDM ($\Lambda$-Cold Dark Matter) and $\omega$CDM models, considering also some specific examples, e.g., Chevallier--Polarsky--Linder, the Chaplygin gas and the Dvali--Gabadadze--Porrati cosmological model. Finally, we consider the cosmological consequences of $f(\mathcal{R})$ and $f(\mathcal{T})$ gravities and their impact on the framework of cosmography. Keeping these considerations in mind, we point out the \emph{model-independent} procedure related to cosmography, showing how to match the series of cosmological observables to the free parameters of each model. \mbox{We critically} discuss the role played by cosmography, as a \emph{selection criterion} to check whether a particular model passes or does not present cosmological constraints. In so doing, we find out cosmological bounds by fitting the luminosity distance expansion of the redshift, $z$, adopting the recent Union 2.1 dataset of supernovae, combined with the baryonic acoustic oscillation and the cosmic microwave background measurements. We perform cosmographic analyses, imposing different priors on the Hubble rate present value. In addition, we compare our results with recent PLANCK  
 limits, showing that the $\Lambda$CDM and $\omega$CDM models seem to be the favorite with respect to other dark energy models. However, we show that cosmographic constraints on $f(\mathcal{R})$ and $f(\mathcal{T})$ cannot discriminate between extensions of General Relativity and dark energy models, leading to a disadvantageous \linebreak degeneracy problem.}


\begin{document}

\section{Introduction}
\label{sec intro}

General Relativity (GR) is today considered a cornerstone of theoretical physics, well supported by a wide number of experiments, in particular, the Solar System ones \cite{citazione1.1,citazione1.2}. All these bounds have been able to fix accurate limits in fairly good agreement with theoretical predictions \cite{citazione2}. Even though Einstein's gravity permits one to describe local gravitational scales with high precision, several observations have showed that the universe is currently undergoing an accelerated phase \cite{SNeIa1.1, SNeIa1.2, SNeIa1.3, SNeIa1.4, SNeIa1.5, SNeIa2, CMBR1.1, CMBR1.2, LSS1.1, LSS1.2, LSS2.1, LSS2.2, Lyalpha}. The former results contradict GR predictions, which provide a decelerated expansion, if one supposes a homogeneous and isotropic universe, dominated by pressureless matter. The contributions coming from, e.g., spatial scalar curvature, radiation, baryons and dark matter (DM) alone are, however, unable to describe the cosmic speed up \cite{citazione3}. As a consequence, the presence of non-interacting matter only (standard and dark) is not enough to drive the accelerated universe dynamics at late times \cite{citazione4}. This shows a possible non-standard origin of the fluid responsible for the acceleration \cite{vnt1.1, vnt1.2, vnt1.3, vnt1.4, vnt1bis}. This fact has been confirmed by the impressive number of high redshift data, through the use of always more accurate distance \linebreak indicators \cite{citazione5}. All those experimental improvements have led to the relevant fact that we are nowadays living in the epoch of {\it Precision Cosmology} \cite{wei}, \emph{i.e.}, one cosmological physics is increasing in the precision of the data. Tighter bounds of the observable universe have been therefore measured, definitively showing that a possible new ingredient \cite{vnt2.1, vnt2.2, vnt2bis}, dubbed dark energy (DE), is actually required in order to accelerate the universe today.

The interpretation of DE in terms of a barotropic fluid addresses the fact that it behaves anti-gravitationally. In order to accelerate the universe, DE must provide a negative pressure, counterbalancing the effects due to Newtonian gravity. To be consistent with current observations, 70$\%$ of the whole universe energy budget seems to be filled by the DE fluid, whose physical interpretation represents a challenge for modern cosmology, since no standard fluid is capable of giving rise to negative pressure \cite{citazione7.1, citazione7.2, citazione7.3}. Furthermore, the thermodynamics of DE is therefore not well understood and represents one of the most intriguing puzzles of cosmology. Even if we adopt barotropic fluid corresponding to a pressure that depends on the energy density only, neither ideal nor real fluids provide the observed negative pressure. In lieu of considering standard fluids, different alternatives have been carried out in the literature, spanning from exotic fluids to anti-gravitational effects due to curvature corrections to quantum gravity effects (see \cite{intermil} and the references therein).

More likely, the simplest explanation of the above problems is provided in the framework of quantum field theories. Indeed, if we include the quantum vacuum energy term within the standard energy momentum tensor, the corresponding energy density shows an upper limit given by the Planck \mbox{mass \cite{citazione8.1, citazione8.2}.} The obtained vacuum energy contribution is different from zero, and it is usually interpreted as \emph{the cosmological constant}, $\Lambda$. In principle, simple calculations show that $\Lambda$ may be responsible for the cosmic speed up \cite{pad}, whereas the cosmological constant fluid behaves as a non-interacting perfect fluid, which does not cluster, as confirmed by several observations. Moreover, the equation of state (EoS) of DE, namely the ratio between the total pressure and density, $\omega\equiv\frac{\mathcal P}{\rho}$, is at any epochs of the universe's evolution $\omega = -1$. This is a consequence of the fact that we consider a constant density and, consequently, a constant pressure. At late times, the DE dominates over matter starting from a redshift known as
\emph{transition redshift} \cite{citazione10.1, citazione10.2}.

It follows that such a vacuum energy cosmological constant \cite{Lambda} could be compatible with early time epoch, and the corresponding model, namely $\Lambda$CDM ($\Lambda$-Cold Dark Matter), also fits the cosmological data at present epoch \cite{wei2}. Despite its experimental guidance, the $\Lambda$CDM model cannot be interpreted as the final cosmological paradigm. The model may represent, rather, a limiting case of a more general theory, which does not reduce necessarily to vacuum energy in terms of a cosmological constant. \mbox{The reason lies} in the fact that the $\Lambda$CDM model is jeopardized by several theoretical shortcomings. In particular, it is plagued by the \emph{coincidence} and \emph{fine-tuning} problems. These two issues today lead to the coincidence between matter density and the cosmological constant order of magnitude and the strong discrepancy between the theoretical and observed magnitudes of $\Lambda$, respectively \cite{iew}. Thus, a plethora of different models have been proposed in order to address these shortcomings \cite{citazione11}.

The simplest extension of $\Lambda$CDM is the $\omega$CDM model \cite{coppa}, which relies on assuming a {\it quintessence} DE term, described by introducing a scalar field, $\phi$ \cite{vnt3}, within GR, providing a constant EoS \cite{blundo}, \emph {i.e.}, $\omega\equiv\frac{\mathcal P}{\rho}=const$. On the one hand, the $\omega$CDM model is a phenomenological coarse-grained extension of the standard cosmological model and naturally reduces to $\Lambda$CDM in the limit $z\ll1$.

On the other hand, the origin of the scalar field and its physical meaning in terms of thermodynamical fluid is not completely clear. In particular, its meaning and behavior at late epochs should be addressed. In turn, we cannot consider the $\omega$CDM model as the definitive paradigm, able to describe the dynamics of DE. An immediate and simple alternative is offered by accounting for an evolving EoS for the DE fluid, originally named {\it varying quintessence} fluid \cite{sa}, which leads to a time-dependent barotropic factor, \emph{i.e.}, $\omega=\omega(z)$.

The aim is to reconstruct the EoS of DE, by demanding that the barotropic factor evolves as the universe expands and comparing it with cosmological data at different stages of the universe's evolution. This prescription is due to the fact that the form of $\omega(z)$ is not defined \emph{a priori}, and so, one may reconstruct its evolution by bounding $\omega(z)$ and its derivatives, through the use of current observations only \cite{ai1,ai2,ai3}.

Several reconstruction methods have been implemented recently. Probably the most relevant approach is offered by the Chevallier--Polarsky--Linder (CPL) parametrization of the EoS \cite{cpl, Linder} in which a Taylor series around the value of the cosmological scale factor $a(t)=a_0=1$ of the barotropic factor is involved, \emph {i.e.}, $\omega(a)=\omega_0+\omega_1(1-a)$.

A relevant advantage of the CPL parameterization is the fact that it predicts an early matter-dominated universe, which agrees with current early time bounds. This represents an improvement of a linear approximation of $\omega(z)$, which diverges as the redshift increases. Rephrasing it, since $a\equiv(1+z)^{-1}$, we have $\omega(z)= \omega_0 + \omega_1 \frac{z}{1+z}$, showing at low and at high redshifts, $\omega(z\rightarrow0)=\omega_0$ and \linebreak $\omega(z\rightarrow\infty)=\omega_0+\omega_1$,~respectively.

However, the degeneracy between $\omega_0$ and $\omega_1$ and the corresponding experimental difficulties of reconstructing the DE evolution plague the CPL parametrization. Hence, even this approach cannot be considered as the standard paradigm, capable of describing DE at all the epochs.

Other viable alternatives do not seem to rely on physical postulates rather than phenomenological approaches, a possibility that concerns a particular hypothesis is offered by extending the standard Einstein--Hilbert action. Under this scheme, introductions of curvature corrections or quantum gravity modifications permit us to reconstruct the DE effects as a natural consequence of a more general theory of gravitation. In this regime, the possibilities of extending GR span from the $f(\mathcal R)$ gravity to Horava--Lifshiz corrections, and so forth \cite{sapore2.1, sapore2.2}. The main disadvantage is, however, the fact that the correct modification is not known \emph{a priori}. Thence, the need for model-independent techniques and experimental tests, able to rule out cosmological models, is probably one of the most important prerogatives of \mbox{modern cosmology.}

In other words, a procedure that \emph{directly} compares cosmological observables in terms of data, instead of cosmological models with data, should guarantee improved constraints, being capable of discarding the models that do not match the corresponding experimental limits. In this regard, the well-known approach of \emph{cosmography} has increased its importance during this time, becoming a relevant model-independent technique to fix cosmological constraints on late time cosmology. In particular, cosmography represents the part of cosmology that studies the universe's dynamics by requiring the validity of the cosmological principle only. Cosmography does not account for any model \emph{a priori}, and it is probably the most powerful technique to derive cosmological bounds directly from data surveys, without passing through the assumption of a cosmological model. \linebreak In so doing, the cosmography prescription consists in expanding the cosmological observables of interest around present time, obtaining Taylor series that can be matched directly with data. The degeneracy problem among cosmological models, which leads to the issue that different models fit data with high precision \cite{sapore3.1, sapore3.2}, is therefore naturally alleviated.

To better understand this statement, let us recall the fact that a single experiment is influenced by a strong model dependence on the cosmological parameters involved in the analysis. This is a consequence of the basic assumption of any cosmological tests, \emph{i.e.}, the model under examination is statistically assumed to be the \emph{best one}.

This leads to the thorny degeneracy of fitting cosmological data, since we cannot distinguish which model is really favored over others.

As a possible landscape, one can intertwine observational tests, \emph{i.e.}, supernovae Ia (SNeIa), baryonic acoustic oscillation (BAO), cosmic microwave background (CMB) and so on, reducing the phase diagram of cosmological data.

In other words, by combining different cosmological tests, it would be possible to reduce the best fit phase space region, healing the degeneracy between cosmological models. For these reasons, cosmography may represent a technique for limiting degeneracy, without necessarily invoking several datasets and without assuming any cosmological model \textit{a priori} \cite{vi1,vi2,komatsu}. The main purpose of this review is to point out the connections between cosmography and cosmological models, showing that it is possible to alleviate the cosmological degeneracy, by adopting the cosmographic model-independent tests. We will discuss in detail the main prescriptions of cosmography, and we will describe the most relevant cosmological models, showing how it is possible to relate such models with cosmographic bounds. In addition, we will take a look at the most common cosmological fits, showing constraints on cosmographic parameters through the use of SNeIa.

This review paper is organized as follows: in Section \ref{sec cosmography}, we discuss cosmography as a selection criterion for determining present time constraints without the need for a cosmological model assumed \textit{a~priori}. In Section \ref{sec issues}, we illustrate the main disadvantages of cosmography and the problems associated with extending the cosmological observables in Taylor series. In Section \ref{sec equazione di stato}, we emphasize how to relate cosmography to the EoS of the universe. In Section \ref{sec DE models}, we review a few cosmological models of particular interest in the framework of modern cosmology. In particular, we summarize the main features of $\Lambda$CDM (Section \ref{sec modello LCDM}), $\omega$CDM (Section \ref{sec modello wCDM}), the CPL parametrization (Section \ref{sec modello CPL}), the Chaplygin gas (Section \ref{sec modello Chap}), the DGP model (Section \ref{sec modello DGP}) and finally, the corresponding phenomenological extension, \emph {i.e.}, the $\alpha$DGP model (Section \ref{sec modello aDGP}). In Section \ref{sec gravita estesa}, we briefly review the so-called Extended Theories of Gravity, focusing on the case of $f(\mathcal R)$ theories (Section \ref{sec f(R)}) and $f(\mathcal T)$ (Section \ref{sec f(T)}). Further, in Section \ref{sec procedure sperimentali}, we introduce the concept of experimental procedures in cosmology, giving light to the observational problem (Section \ref{sec problemi osservativi}), BAO, SNeIa and CMB measurements. Finally, Section \ref{sec conclusioni} is devoted to conclusions and the perspectives of this review.

\section{Cosmography: A Selection Criterion for Cosmological Models} \label{sec cosmography}

 We describe in detail the role played by cosmography and its consequences on observational cosmology.
Cosmography represents a selection criterion to discriminate which model behaves better than others in comparison with cosmological data~\cite{cv1}. In fact, all numerical tests depend on the choice of the particular cosmological model under examination, leading to a strong degeneracy problem. In order to alleviate the degeneracy between cosmological models, one needs to introduce \emph{model-independent} procedures for constraining cosmological scenarios, discriminating the statistically favorite paradigm.
An intriguing landscape to the degeneracy problem is to take into account the cosmological quantities that can be fitted without the need for postulating a model \textit{a priori}. Cosmography was first widely discussed by Weinberg \cite{wei} and then extended in~\cite{cv3,orlandoluongo}. The \emph{core} of cosmography is to assume the validity of the cosmological principle only, without any further assumptions on Einstein's equations. \mbox{In other words,} once the Friedmann--Robertson--Walker (FRW) metric is involved, cosmography teaches us how to get bounds on the observable universe, through direct measurements of the observable expansions in terms of $a(t)$. Therefore, if we write the FRW metric as:
\begin{equation}\label{frpoaaaaaa}
ds^2=dt^2-a(t)^2 \left( \frac{dr^2}{1-kr^2}+r^2\sin^{2}\theta d\phi^2+r^2d\theta^2 \right) \,
\end{equation}
with $a(t)$ the scale factor and $k$ the standard spatial curvature, then cosmography permits us to infer how much DE or alternative components are required to guarantee current observations, without postulating any cosmological model at the beginning of our analyses. The idea is to relate the cosmographic expansions to the free parameters of a given model.

Hence, one can appraise which models behave fairly well and which are disfavored as a consequence of not satisfying the basic demands introduced by cosmography. Usually, we refer to cosmography as a part of cosmology that tries to infer kinematical quantities by keeping the FRW geometry only. Rephrasing it, the framework of cosmography is represented by assuming Taylor expansions of the scale parameter as the basic postulate. In so doing, standard Hubble law becomes a Taylor series expansion for low $z$. Starting from the Hubble law $v$ = $\mathcal{H}_0 D$, it is easy to obtain the \mbox{following expression}:
\begin{equation}
\frac{a_0}{a(t)} = 1 + \mathcal{H}_0 D\,
\end{equation}
where $D$ is the causal distance traveled by photons from an emitter to an observer. Since $1 + z = 1 + v/c$, it follows:
\begin{equation}\label{mmcmcdncnncnc}
1 + z = \frac{a_0}{a(t)}\,
\end{equation}
which is the redshift definition in terms of the scale factor $a(t)$. Afterwards, we can evaluate a Taylor series expansion of this quantity, obtaining an expression of the redshift in terms of the physical distance, $z(D)$. These descriptions represent the basic ingredient to expand the scale factor into a series, yielding:
\begin{eqnarray}
a(t) \approx a_0\Bigg[ 1 + \frac{da}{dt}\Big|_{t_0} (t-t_0) + \frac{1\,}{2!}\frac{d^2a}{dt^2}\Big|_{t_0} (t-t_0)^2 + \frac{1\,}{3!} \frac{d^3a}{dt^3}\Big|_{t_0} (t-t_0)^3 + \frac{1\,}{4!} \frac{d^4a}{dt^4}\Big|_{t_0} (t-t_0)^4 + \ldots \Bigg]\,
\label{eq:expa}
\end{eqnarray}
where we conventionally truncated the series at the fourth order in $\Delta t\equiv t-t_0$. Here, we assume that $t-t_0>0$, in order to get the causality of cosmological observations. By simply assuming that the constant, $a_0$, \emph {i.e.}, the scale factor value today, is $a_0=1$, we have:
\begin{equation}\label{serie1a}
a(t) \approx 1 + \mathcal{H}_0 \Delta t - \frac{q_0}{2} \mathcal{H}_0^2 \Delta t^2 +
\frac{j_0}{6} \mathcal{H}_0^3 \Delta t^3 +
 \frac{s_0}{24} \mathcal{H}_0^4 \Delta t^4 + \dots\,
\end{equation}
with the definition of the cosmographic series (CS) as:
\begin{equation}\label{pinza}
 \mathcal{H} \equiv \frac{1}{a} \frac{da}{dt}\,, \quad q \equiv -\frac{1}{a\mathcal{H}^2} \frac{d^2a}{dt^2}\,, \quad j \equiv \frac{1}{a\mathcal{H}^3} \frac{d^3a}{dt^3}\,, \quad s \equiv \frac{1}{a\mathcal{H}^4} \frac{d^4a}{dt^4}\,
\end{equation}

The Hubble rate represents the first derivative with respect to cosmic time of the logarithm of the scale factor $a$, whereas the acceleration parameter, $q$, indicates if the universe is currently accelerated or not. A currently accelerating universe provides $-1 \leq q_0 \leq 0$, where the subscript, $0$, points out that $q$ \linebreak (and all the cosmographic parameters) are evaluated at present time. The variation of acceleration, namely the jerk parameter, $j$, if positive, would indicate that $q$ changed sign in the past, at the transition redshift, corresponding to $q=0$. Finally, $s_0$ indicates if $j$ changed sign as the universe expands. \mbox{If negative}, the jerk parameter remains with the same sign of its present measurable value.

After straightforward calculations, we could obtain an expression of the luminosity distance, $d_L$, expanded in series around $z \simeq 0$. In particular, considering the expression for the luminosity distance and assuming the causal distance:
\begin{equation}\label{distanza}
D =\int_{0}^{z}\frac{d\psi}{\mathcal{H}(\psi)}\,
\end{equation}
with the definition of $d_L$ given by:
\begin{equation}\label{distanza}
d_L (z) =(1+z)\int_{0}^{z}\frac{d\psi}{\mathcal{H}(\psi)}\,
\end{equation}
we have:
\begin{eqnarray}\label{zeriez}
d_L(z) &\approx& d_H\,z\Bigg\{ 1 + {\left[1-q_0\right]\over2} {z}-{1\over6}\left[j_0+ \Omega_0-q_0-3q_0^2\right] z^2+ \nonumber \\
 && +z^3 \Bigl( \frac{1}{12} + \frac{5 j_0}{24} - \frac{q_0}{12} + \frac{5 j_0 q_0}{12} - \frac{5 q_0^2}{8} - \frac{5 q_0^3}{8} + \frac{s_0}{24} \Bigr)\Bigg\} \,
\end{eqnarray}
where $d_H = 1/\mathcal{H}_0$ is the Hubble radius and $\Omega_0$ the total energy of the universe, which can be related to the spatial scalar curvature, $\Omega_{k}\equiv\frac{k}{a^2}$, in terms of $\Omega_k=\Omega_0-1$. As we will clarify later, the closure relation, \emph{i.e.}, $\Omega_k=0$, would provide a total density fixed to $\Omega_0=1$.

For the sake of clarity, we want to underline the fact that cosmography can be extended in the framework of an inhomogeneous universe. In that case, extensions of the CS are required in order to \emph{define} a new set of cosmographic parameters. The role played by the acceleration parameter, the jerk parameter, and so forth, must be inferred by starting from a different metric. An example of inhomogeneous space time is provided by the Lemaitre--Tolman--Bondi metric \cite{dqc1}. However, a self-consistent scheme for inhomogeneous cosmography is not, so far, completely understood (for additional details, see \cite{dqc2}). Inhomogeneous cosmographic investigations are extremely important in order to measure possible departures from the standard homogeneous and isotropic universe. However, in this review, we do not treat the case of inhomogeneous cosmography extensively.

\section{Issues Related to Cosmography} \label{sec issues}

Although cosmography relates cosmological parameters to power series expansions, permitting one to directly measure cosmological observables, the problem of assuming a precise model-independent procedure depends on $\Omega_0$. In fact, the measurement of $j_0$ cannot be performed alone. This comes from the fact that one can only measure the sum $j_0+\Omega_0$, instead of $j_0$. The former prescription entails some other difficulties on the numerical procedure, since Taylor expansions cannot alleviate the degeneracy between $j_0$ and $\Omega_0$. In other words, to reduce this problem, one needs geometrical ways to constrain the universe scalar curvature, \emph {i.e.}, finding out limits that permit one to fix scalar curvature. In particular, if the DE term would be completely time independent, the bounds inferred from CMB measurements indicate a compatible spatially flat universe. On the contrary, if the EoS of DE is a function of the cosmic time, limits on $\Omega_k$ are not so restrictive, and it is not completely clear if space curvature is really negligible. However, at present times, the universe can be fairly well approximated through a vanishing scalar curvature, and so, we will not lose generality imposing hereafter $\Omega_k=0$. In turn, cosmography becomes, therefore, a model-independent procedure to constrain cosmological data. On the one hand, cosmography constrains cosmological models without the need for a particular theoretical framework, but on the other hand it, approximates the original function at a certain order of the truncated series. Drawbacks associated with those approximations are therefore involved. In fact, the exact expansion of cosmological parameters will clearly be different from a truncated series with an error as small as the series tends to higher orders. If one truncates the series at small orders, corresponding errors would propagate into the analyses, considerably complicating the resulting constraints. Even if the analyses are plagued by a truncated series problem, it is desirable to keep the number of fitting parameters as low as possible, in order to reduce the broadening of the posterior distributions for each parameter, derived from extending the parameter space. In other words, a low order of the Taylor expansion would propagate numerical departures within the error bars of the cosmographic parameter. However, a high order may increase the phase space, enlarging the allowed statistical regions of cosmic parameters. To alleviate these shortcomings, one can consider different datasets, combining cosmological data with more than one survey compilation. Afterwards, one may truncate at an intermediate order of the Taylor series all the observable quantities of interest in order to reduce systematics on measurements.

In addition to the above problems, although cosmography represents a viable technique to perform model-independent fits, the numerical procedure of expansion around our time is plagued by problems related to divergences in the Taylor series of the CS. The CS may diverge at $z \gtrsim 1$, since we are expanding around $z\sim0$, as a consequence of the particular convergence radius, associated with each expansion. In other words, when one exceeds the approximation of low redshift, problems due to convergence may occur, and so, finite truncations could give systematics on the measurements of cosmological quantities, providing possibly misleading results.


One well consolidated way to escape the convergence problem is represented by re-parameterizing the redshift variable, through the use of auxiliary variables, circumscribed in a low redshift domain, in which any possible parameterizations of the redshift, $z$, namely $y_i$, may be limited to $y_i<1$. An example is the following well-known variable:
\begin{equation}
y_1 = {z\over1+z}
\end{equation}
which has the advantage of converging for past and future times, respectively:
\begin{equation}
z \in (0,\infty); \qquad y\in (0,1)\,; \qquad
z \in (-1,0); \qquad y\in (-\infty,0)\,
\end{equation}
one can rewrite the luminosity distance in terms of the auxiliary variable, expanding in a Taylor series around it. The corresponding discrepancies between the two procedures do not influence the systematics of the cosmographic analysis, and so, rewriting the luminosity distance of Equation (\ref{zeriez}) in terms of $y_1$, we have
\cite{caps}:
\begin{eqnarray}\label{mah}
d_L(y_1) \approx && d_H \left\{y_1-\frac{1}{2}(q_0-3)y_1^2+\frac{1}{6}\left[12-5q_0+3q^2_0-(j_0+\Omega_0)\right]y_1^3+\frac{1}{24} \left(60-7j_0 + \right. \right.\nonumber
 \\ && \left.\left.-10\Omega_0-32q_0+10q_0j_0+6q_0\Omega_0+21q^2_0-15q^3_0+s_0\right) y_1^4+\mathcal{O}(y_1^5) 
 \vphantom{ \left( \frac{c}{\mathcal{H}_0} \right) } 
 \right\}\,
\end{eqnarray}
where $\Omega_0$ is the net energy density of the universe. On the other side, other interesting variables have been introduced \cite{orlandoluongo}:
\begin{eqnarray}\label{ilbosco}
y_2=\arctan{\left(\frac{z}{z+1}\right)}\,,\qquad y_3=\frac{z}{1+z^2}\,, \qquad y_4&=&\arctan{z}\,
\end{eqnarray}
whose limits are, for $z \in [0,\infty]: y_2\in [0,\frac{\pi}{4}],
y_3\in [0,0], y_4\in [0,\frac{\pi}{2}]$ and for $z \in [-1,0]: \linebreak y_2\in
[-\frac{\pi}{2}, 0], y_3\in [-\frac{1}{2},0], y_4\in
[-\frac{\pi}{4},0]$. Such variables respect the condition of finite intervals in which the redshift is scaled to values that do not allow bad convergence and divergence in the \mbox{Taylor expansions.}

Some interesting steps to contrive viable redshift parameterizations are based on the reconstruction of the redshift variable and on the invertibility with respect to $z$. In particular, one may assume that the luminosity distance curve should not behave too steeply in the interval $z<1$, providing a smooth profile, without flexes. Rephrasing it mathematically, the curve should be one-to-one invertible, in terms of the auxiliary variables, providing tighter intervals for redshifts. For the sake of completeness, the corresponding luminosity distances are rewritten as follows:

\begin{eqnarray}
d_L(y_2)& \approx & d_H \left[ y_2 + y_2^2 \left(\frac{3}{2} - \frac{q_0}{2} \right)
 + y_2^3 \left(\frac{13}{6} -\frac{j_0}{6} - \frac{5q_0}{6} + \frac{q_0^2}{2} \right) + \right. \nonumber\\ && \left.+ y_2^4 \left( \frac{37}{12} - \frac{7 j_0}{24} - \frac{17 q_0}{12} + \frac{5 j_0 q_0}{12}
 - \frac{7 q_0^2}{8} - \frac{5 q_0^3}{8} + \frac{s_0}{24} \right) + \ldots \right]\,
 \end{eqnarray}
\begin{eqnarray}
d_L(y_3) & \approx & d_H \left[ y_3 + y_3^2 \left(\frac{1}{2} - \frac{q_0}{2} \right)
 + y_3^3 \left(\frac{5}{6} -\frac{j_0}{6} + \frac{q_0}{6} + \frac{q_0^2}{2} \right) + \right. \nonumber\\ && \left.+ y_3^4 \left( \frac{13}{12} + \frac{5 j_0}{24} - \frac{13 q_0}{12} + \frac{5 j_0 q_0}{12} - \frac{5 q_0^2}{8} - \frac{5 q_0^3}{8} + \frac{s_0}{24} \right) + \ldots\right]\,
 \end{eqnarray}
 \begin{eqnarray}
d_L(y_4)& \approx & d_H \left[ y_4 + y_4^2 \left(\frac{1}{2} - \frac{q_0}{2} \right)
 + y_4^3 \left(\frac{1}{6} -\frac{j_0}{6} + \frac{q_0}{6} + \frac{q_0^2}{2} \right) + \right. \nonumber\\ && \left.+ y_4^4 \left( \frac{5}{12} + \frac{5 j_0}{24} - \frac{5 q_0}{12} + \frac{5 j_0 q_0}{12} - \frac{5 q_0^2}{8} - \frac{5 q_0^3}{8} + \frac{s_0}{24} \right) + \ldots \right]\,
 \end{eqnarray}
respectively, for $y_2,y_3$ and $y_4$. Note again that we adopted $\Omega_0=1$ in the above expansions.

The need for accurate cosmographic analyses relies on another important fact. Indeed, we notice that DE can also be interpreted not in terms of a mysterious substance. The interpretation of DE in terms of a mysterious fluid is a consequence of assuming a further fluid within Einstein's field equations, \linebreak to drive the observed universe speed up. On the contrary, carrying out accurate cosmographic analyses would permit cosmologists to denote DE through possible different interpretations. This philosophy can open new insights for discovering the physics behind DE (for example, anthropic principle, holographic principle, accelerating frames, and so on \cite{rover}). In other words, since cosmography could discriminate the real nature of DE, improved cosmographic analyses would refine our theoretical knowledge on the observable universe, as well.

\section{Cosmography and the Equation of State of the Universe} \label{sec equazione di stato}

The CS represents a resource for determining cosmological bounds without the need for a cosmological model postulated \textit{a priori}. The subscript ``$0$'' indicates that the CS is evaluated at redshift $z=0$, which represents our time. It is useful to combine the CS among themselves, obtaining a combination at a generic time, $t$:
\begin{subequations}\label{ciao}
\begin{align}
q(t)&=-\frac{\dot{\mathcal{H}}}{\mathcal{H}^2} -1\,\\
j(t)&=\frac{\ddot{\mathcal{H}}}{\mathcal{H}^3}-3q-2\,\\
s(t)&=\frac{\dddot{\mathcal{H}}}{\mathcal{H}^4}+4j+3q(q+4)+6\,
\end{align}
\end{subequations}
where we assumed $q,j$ and $s$ in terms of the Hubble rate, \emph{i.e.}, $\mathcal{H}\equiv \frac{d}{dt}\ln a$. The dots indicate the derivatives with respect to the cosmic time. Analogously, one can expand another important quantity of cosmological interest, \emph {i.e.}, the total pressure as a function of the whole energy density \mbox{$ \mathcal{P} = \mathcal{P} (\rho)$}, as:
\begin{equation}\label{basta}
\mathcal{P} \approx \mathcal{P}_0 + \kappa_0^{(1)} (\rho-\rho_0) + \frac{1}{2}\kappa^{(2)}_{0}(\rho-\rho_0)^2 + \ldots\,
\end{equation}
with $\kappa_{0}^{(1)}\equiv\frac{d \mathcal{P}}{d\rho}\Big|_{0}$, $\kappa^{(n)}_{0}\equiv\frac{d^n \mathcal{P} }{d\rho^n}\Big|_{0}$ evaluated at $z=0$ and truncated to $n=2$. Simple algebra leads to:
\begin{subequations} \label{eq:pressureandD}
\begin{align}
 \mathcal{P} &= \frac{1}{3}\mathcal{H}^2 \left( 2q - 1\right)\,\label{eq:pressure} \\
\frac{d \mathcal{P} }{dt} & = \frac{2}{3} \mathcal{H}^3 \left(1 - j\right)\, \label{eq:pressure1} \\
\frac{d^2 \mathcal{P} }{dt^2} & = \frac{2}{3} \mathcal{H}^4 \left(j- 3 q - s -3 \right)\,
 \label{eq:pressure2}
\end{align}
\end{subequations}
where we used the fact that the total density of the universe is a function of the redshift, $z$, \emph {i.e.}, $\rho=\rho(z)$.

Simple calculations permit one to relate the EoS of the universe in terms of cosmographic coefficients. We have:
\begin{equation}\label{equazionecosmografica}
\omega = \frac{\mathcal{P}}{\rho} = \frac{2q-1}{3}\,
\end{equation}
which represents the \emph{total} EoS of all species that enter the standard energy momentum tensor. \mbox{At present time}, Equation (\ref{equazionecosmografica}) can be expressed in terms of standard matter and DE. We therefore obtain:
\begin{equation}\label{wtotmattde}
\omega=\frac{\mathcal{P}_{DE}}{\rho_m+\rho_{DE}}\,
\end{equation}
where we have used the fact that $\rho=\rho_m+\rho_{DE}$ and the matter pressure vanishes $\mathcal{P}_m=0$.

With each species, we can associate the corresponding EoS, which is commonly different from the one of Equation (\ref{wtotmattde}).

We briefly report below some examples of the EoS of single species:
\begin{eqnarray}
 \omega_m=0\,, \qquad \omega_k=-\frac{2}{3}\,, \qquad \omega_r = \frac{1}{3}\,
\end{eqnarray}
respectively, for standard pressureless matter (baryons and CDM), scalar curvature and radiation.

In general, the expression of the EoS of DE reads:
\begin{equation}\label{dewwwwwww}
\omega_{DE}=\frac{ \mathcal{P}_{DE} }{\rho-\rho_m}\,
\end{equation}
in which we neglected radiation and scalar curvature, to be consistent with the current bounds on the observable universe.

Note that, generally, $\omega_{DE}\neq\omega$, and so, the corresponding CS is different if the universe is characterized by one, two or more fluids and, further, that $\rho>\rho_{DE}$.
%
%

\section{Dark Energy Cosmological Models} \label{sec DE models}

We discussed the role of cosmography in the present status of cosmology. We described how to relate the derivatives of the pressure and EoS in terms of the CS. In this section, we wonder whether it is possible to relate the total matter density of a particular cosmological model in terms of the CS. \mbox{This prescription} is actually important, since the allowed limits on matter density can be inferred from measuring the CS. A model that does not confirm cosmographic bounds is disfavored over the ones that pass those constraints. This methodology is sometimes refereed to as \emph{the cosmographic test}, and it is similar to the statefinder diagnostic. To perform a cosmographic test, we briefly sketch some cosmological scenarios among the most relevant ones. We can therefore investigate the numerical predictions of any model through the use of cosmography.

As already stressed in the introduction, we focus on the simplest assumption of an homogeneous and isotropic universe. The corresponding Friedmann equations read:
\begin{equation}\label{constr}
 \mathcal{H}^2\equiv\left(\frac{\dot a}{a}\right) = \frac{8\pi G}{3}\rho-\frac{k}{3} \,,
\qquad \dot{\mathcal{H}} +\mathcal{H}^2 = -\frac{4\pi G}{3} \left( 3\mathcal{P} + \rho \right) \,
 \end{equation}
which have been written in general, leaving the scalar curvature $k$. The energy-momentum conservation for a barotropic fluid, \emph {i.e.}:
$\partial^\mu T_{\mu\nu}=0$ provides,
\begin{equation}\label{lej}
\dot{\rho} +3 \mathcal{H} \left( \mathcal{P} + \rho \right) =0 \,
\end{equation}
and with the definition of the cosmic time, in terms of $z$, \emph {i.e.}:
\begin{equation}\label{vvvvvvnnnnnnnnn}
\frac{dz}{dt}=-(1+z)\mathcal{H}(z)\,
\end{equation}
we can recast a convenient expression for the continuity equation, rewriting Equation (\ref{lej}) as follows:
\begin{equation}\label{lej2}
\frac{d\rho}{dz}=3\left(\frac{ \mathcal{P} +\rho}{1+z}\right)\,
\end{equation}
From Equation (\ref{lej2}), it is easy to probe:
\begin{equation}\label{ronzo}
 \rho \propto \exp\left[{3\int{\frac{1 + \omega(z)}{1+z}}dz}\right]\,
\end{equation}
which teaches us how to relate the total energy density to the barotropic factor, $\omega(z)$. Depending on the particular cosmological model or, alternatively, on $\omega(z)$, one gets different evolutions of the total \mbox{density, $\rho$.}

For barotropic fluids, $\omega$ degenerates with the spatial curvature density, $\Omega_k$, and as previously stressed, one should include $\Omega_k$ as an additional parameter. Current data are not accurate enough to distinguish between $\omega(z)$ and $\Omega_k$ alone, owing to a strong degeneracy problem between them.

As discussed, this condition is transferred to the measurements of the jerk parameter, $j$, which cannot be constrained alone if the scalar curvature is not fixed by some geometrical bound. Here, for simplicity, we assume a negligible spatial curvature, $\Omega_k$ \cite{coppa}, as already discussed in previous sections.

To relate cosmography with cosmological models, we take into account the following paradigms:
\newpage
\begin{enumerate}
\item The $\Lambda$CDM model;
\item DE models with a constant equation of state, derived from a scalar field coupling with curvature ($\omega$CDM or quintessence);
\item DE models with EoS parameterized in terms of the power of $a(t)$ (e.g., the CPL parametrization);
\item DE models interacting with CDM (e.g., the \textit{Chaplygin gas});
\item DE from quantum effects (e.g., the Dvali--Gabadadze--Porrati (DGP) model and its \linebreak phenomenological extension);
\item $f(\mathcal R)$-gravity models;
\item $f(\mathcal T)$-gravity models.
\end{enumerate}

In the upcoming sections, we are going to briefly review each of them, before discussing the connections among them and cosmography.

%
\section{The $\Lambda$CDM Model} \label{sec modello LCDM}

The cosmological constant was originally achieved by Einstein himself, in order to reproduce a static universe. After the discovery of a non-static universe, the cosmological constant was discarded, due to its original bad interpretation. Recently, its reintroduction has been proposed, in order to guarantee that the universe accelerates. The physical meaning of $\Lambda$ is, however, at most different from the one originally proposed by Einstein. In particular, given the Einstein tensor $G_{\mu\nu}=\mathcal R_{\mu\nu}-\frac{1}{2}g_{\mu\nu}\mathcal R$ and the corresponding energy momentum tensor, $T^{\mu \nu}$:
\begin{equation}\label{mq}
T^{\alpha \beta}=(\rho +\mathcal{P})u^{\alpha} u^{\beta} - \mathcal{P} g^{\alpha \beta}\,
\end{equation}
both satisfying the Bianchi identities, it would be natural to include a constant term, $\Lambda$, which makes the Einstein
equations the most general possible. Indeed, an integration constant does not influence the validity of the Bianchi identities, and in the context of standard GR, quantum field theory interprets the cosmological constant as a vacuum energy contributor, which enters the right side of Einstein's equations. In a homogeneous and isotropic background, the Friedmann equations are rewritten analogously to Equation (\ref{constr}), by only replacing $\mathcal{P}$ and $\rho$ with:
\begin{eqnarray}\label{perho}
\rho \rightarrow \rho + \frac{\Lambda}{8\pi G}\,,\qquad
\mathcal{P} \rightarrow \mathcal{P} -\frac{\Lambda}{8\pi G}\,
\end{eqnarray}
Despite its theoretical interpretation, the meaning of $\Lambda$ is nowadays the object of strong debates. Even if particle physics
interprets $\Lambda$ as a vacuum energy density, the role of $\Lambda$ is plagued by a severe fine-tuning problem. Particularly, the vacuum energy density is naively evaluated by the sum of the zero-point energies of quantum fields with mass $m$ by:
\begin{equation}\label{massam}
\rho_{\rm vac} =\frac{1}{4\pi^2} \int_{0}^\infty d \kappa \, \kappa^2 \sqrt{\kappa^2+m^2}\,
\end{equation}
which exhibits an ultraviolet divergence, overcoming by assuming a cut-off scale, $\kappa_{\rm max}$, proportional to the Planck mass, as already discussed in the Introduction.

As a consequence:
\begin{equation}\label{conseqgueaz}
\rho_{\rm vac} \approx
\frac{\kappa_{\rm max}^4}{16\pi^2}\,
\end{equation}
which is about $10^{120}$ orders of magnitude larger than the observational value measured by \linebreak cosmological observations.

This shortcoming is known in the literature as \emph{fine-tuning} problem. In addition, a coincidence problem plagues the model, \emph {i.e.}, the present value of $\Lambda$ density is comparable with matter density $\Omega_m$ today.

If we define the cosmological constant density as $\Omega_\Lambda$, we find out:
\begin{equation}\label{nsgaehfkehef}
\frac{\Omega_{\Lambda}}{\Omega_m}\approx 2.7\,
\end{equation}
which appears as a strange causality, since $\Omega_\Lambda$ does not evolve as the universe expands. Hence, it is not clear why we would live exactly in an era in which the two orders are so close to each other.

By considering Equations (\ref{constr}) and (\ref{perho}), the normalized Hubble rate, $ \mathcal{E} (z)\equiv\frac{\mathcal{H}(z)}{\mathcal{H}_0}$, easily reads:
\begin{equation}\label{nncivo}
\mathcal{E}(z)=\sqrt{\Omega_{m}(z)+\Omega_\Lambda} \,
\end{equation}
\noindent where, hereafter, $\Omega_m(z) \equiv \Omega_{m}(1+z)^3 $ and, to guarantee $\mathcal{E}(z=0)=1$, $\Omega_\Lambda=1-\Omega_m$. The $\Lambda$CDM model depends on one-parameter only, \emph{i.e.}, $\Omega_m$, and provides the following expressions for the CS:
\begin{equation}\label{greta}
q_0= \frac{3}{2}\Omega_m -1 \,
\end{equation}
\begin{equation}\label{j054}
j_0= 1 \,
\end{equation}
and:
\begin{equation}\label{s0l}
s_0= 1 - \frac{9}{2} \Omega_m \,
\end{equation}
Inverting Equation (\ref{greta}), we get:
\begin{equation}\label{inversione1}
\Omega_{m,q_0}=\frac{2}{3}\left(1 + q_0 \right) \,
\end{equation}
which represents the matter density as a function of the acceleration parameter. It is easy to show that if a model is one parameter dependent, we expect that it depends on $q_0$ only. If a model depends on more than one parameter, then we need to relate its $\Omega_m$ even to $j_0$ and $s_0$.

\section{The Quintessence Model} \label{sec modello wCDM}

In analogy to the $\Lambda$CDM model, we wonder whether the origin of the negative barotropic factor may be derived from prime principles. Probably the simplest approach, generalizing the $\Lambda$CDM model, deals with assuming that an unknown component, namely quintessence, is derived, for example, from a \mbox{slowly-rolling} dynamical scalar field with a non-canonical kinetic term in the Lagrangian. The corresponding paradigm predicts a constant EoS, in which the barotropic factor is consistent with modern observations. For the sake of clarity, the cosmological model is named $\omega$CDM, due to its analogy with the $\Lambda$CDM paradigm. Here, the role of the cosmological constant is replaced by a constant barotropic factor. The normalized Hubble rate, $\mathcal{E}(z)$, then reads:
\begin{equation}\label{XCDM}
\mathcal{E}(z)=\sqrt{\Omega_{m}(z)+\Omega_Q(1+z)^{3(1+ \omega )}} \,
\end{equation}
The $\omega$CDM model is a two-parameters model, $\Omega_m$ and $\omega$, and $\Omega_Q=1-\Omega_m$. The $\omega$CDM model leads to the following CS:
\begin{equation}\label{q0l}
q_0= \frac{1}{2} [1 - 3\omega ( \Omega_m - 1)] \,
\end{equation}
\begin{equation}\label{j0l}
j_0= \frac{1}{2} [2-9 \omega (\Omega_m - 1) (1 + \omega )] \,
\end{equation}
and:
\begin{eqnarray}\label{s0l}
s_0= \frac{1}{4} \{-38-9 \omega (\Omega_m - 1) \{ -9 + \omega [-16 -9 \omega +3 \Omega_m (1+\omega)] \} \} \,
\end{eqnarray}
By inverting the coefficients, we have:
\begin{equation}\label{inversione3000}
\Omega_{m,q_0}=\frac{(1+3\omega) -2q_0}{3\omega} \,
\end{equation}
and:
\begin{equation}\label{invghhhh}
\Omega_{m,j_0}=\frac{\left[ 2(1-j_0) +9\omega(1+\omega) \right] }{9\omega(1+\omega)}\,.
\end{equation}
Alternatively, the $\omega$ parameter reads:
\begin{equation}\label{inversione6000}
\omega=-\frac{1 + 2j_0-6q_0}{3\left(1-2q_0\right)} \,
\end{equation}
and so, combining Equations (\ref{inversione3000}) and (\ref{inversione6000}), we have:
\begin{equation}
 \Omega_{m,q_0,j_0} = 2 \cdot \frac{ j_0 - q_0 ( 1 + 2 q_0 ) }{1 + 2 j_0 - 6 q_0 }\,
\end{equation}
which represents the value of $\Omega_m$ in terms of $q_0$ and $j_0$, as expected, since $\omega$CDM depends on \mbox{two parameters.}

\section{The Chevalier--Polarsky--Linder Parameterization} \label{sec modello CPL}

Differently from the two mentioned approaches, we may include a rolling scalar field, $\phi$, in order to
establish a direct parametrization of the EoS of DE.

In general, to characterize the DE density, one needs to know the form of $\omega=\omega(z)$, obtaining the corresponding normalized Hubble rate:
\begin{equation}\label{Hparaz}
\mathcal{E}(z)= \sqrt{\Omega_{m}(z) + \left(1- \Omega_{m}\right) f(z)} \,
\end{equation}

\noindent where $f(z)$ is a generic function of $z$, so providing the evolving DE term depending on the redshift, $z$. We have:
\begin{equation}\label{yh51.9}
f(z) =\exp
\left[3 \int_{0}^z \frac{1+\omega(\tilde{z})} {1+\tilde{z}} d
\tilde{z}\right] \,
\end{equation}
If the evolution of $\omega(z)$ is the result
of a rolling scalar field, a simple choice is to parameterize $\omega(z)$ in terms of a Taylor expansion around $z=0$:
\begin{equation}\label{mbfffffffffffff}
\omega(z)=\sum_{n=0}^{\infty} \omega_n z^n\,
\end{equation}
where the coefficients, $\omega_n$, are the derivatives of $\omega$ evaluated at $z=0$.

Assuming a direct Taylor expansion leads to the problem of truncating the series and may provide significant discrepancies with observations at higher redshifts. This drawback can be alleviated by matching Taylor expansions directly to data. Many approaches have been developed in the literature, and here, we discuss the CPL parametrization, given by:
\begin{equation}
\label{cpl1}
\omega(z)=\omega_0+\omega_1\frac{z}{1+z}\,
\end{equation}
where $\omega_0$ and $\omega_1$ are constants. One can interpret the CPL parametrization as a first order Taylor series in the powers of $a(t)$, around $a=1$.
The corresponding normalized Hubble rate is:
\begin{equation}\label{cpl2}
\mathcal{E}(z)=\sqrt{\Omega_{m}(z)+\Omega_{DE}(1+z)^{3(1+\omega_0+\omega_1)}\exp\left(-\frac{3\omega_1
z}{1+z}\right)} \,
\end{equation}
where $\Omega_{DE}=1-\Omega_m$.

In this case, the model depends on three parameters, $\Omega_m,\omega_0,\omega_1$, whereas the CS reads:
\begin{equation}\label{q0l}
q_0=\frac{1}{2} [1 - 3\omega_0 (\Omega_m - 1)] \,
\end{equation}
\begin{equation}\label{j0l}
j_0=\frac{1}{2} [2 - 9 \omega_0 (\Omega_m -1) (1 + \omega_0) + 3 \omega_1 -3 \Omega_m \omega_1 ] \,
\end{equation}
and:
\begin{eqnarray}\label{s0l}
s_0&=&\frac{1}{4} \{-14 - 9 \omega_0 (\Omega_m - 1) \{ -9 + \omega_0 [-16 - 9 \omega_0 + 3 \Omega_m (1 + \omega_0) ] \} + \nonumber \\
 && - 33 \omega_1 + 3 \omega_1 [11 \Omega_m -3 \omega_0 (\Omega_m - 7) (\Omega_m - 1) ] \} \,
\end{eqnarray}
and so:
\begin{equation}\label{inversione7}
\Omega_{m,q_0}=\frac{1 -2q_0+3\omega_0 }{3\omega_0} \,
\end{equation}
\begin{equation}\label{inversione8}
\Omega_{m,j_0}=\frac{ 2 +9\omega_0 +9 \omega_{0}^{2}+3\omega_1 -2j_0}{3\left(3\omega_0+3\omega_{0}^{2}+\omega_1\right)} \,
\end{equation}
while:
\begin{equation} \label{inversione10}
 \begin{split}
 \omega_0 &= \frac{1 -2q_0}{3\left( \Omega_m - 1\right)} \, \\
\omega_1 &= \frac{-2 \left(2 q_0 + 1\right) q_0 -\Omega_m \left(1 - 6q_0\right)+2j_0\left(1 -\Omega_m\right)}{3\left(\Omega_m-1\right)^2} \,
 \end{split}
\end{equation}
Finally:

\begin{equation}
 \Omega_{q_0, j_0, \omega_1} = \frac{-1 - 2 j_0 + 6 q_0 + 6 \omega_1 + \sqrt{(1+2 j_0 - 6 q_0)^2 - 12 (1-2 q_0)^2 \omega_1}}{6 \omega_1} \,
\end{equation}
\begin{eqnarray}
 \Omega_{m,q_0,j_0,s_0} &= & \Bigl[ -1+7 j_0+2 q_0 + 8 j_0 q_0 - 16 q_0^2 + 2 s_0 + \nonumber \\
 && + (1- 2 q_0 ) \sqrt{1+9 j_0^2 + 2 j_0 (4 q_0^2 -6 q_0 -7)-4 s_0+8 q_0 (4 q_0+s_0)} \hspace{0.1cm} \Bigr] \times \\
 && \times \{2 [s_0 -4 q_0+j_0 (5+q_0)] \}^{-1} \nonumber \,
\end{eqnarray}

In order to constrain $\Omega_m$, one needs to know with high precision $q_0,j_0$ and $s_0$, since the CPL parametrization is a three-parameter-dependent model. Hence, we expect a broadening of systematics in evaluating the coefficients, since the numerical values of the CS, as we will see later, are not well constrained by present data.

\section{The Chaplygin Gas} \label{sec modello Chap}

Now, we fix our attention on the \textit{Chaplygin gas} model. Chaplygin first introduced his equation for a perfect fluid in 1904 \cite{Chaplygin}, proposing as pressure:

\begin{equation} \label{chap}
\mathcal{P} = - \frac{A}{\rho} \,
\end{equation}
where $A$ is a positive constant. In particular, $\rho$ and $\mathcal{P}$ are the (total) density and pressure in a comoving reference domain, where $\rho > 0$. Chaplygin treated the particles of its fluid as dust, obtaining a vanishing pressure contribution to standard matter. By combining Equations (\ref{lej2}) and (\ref{chap}), we get:
\begin{equation} \label{rho chap}
 \rho(a) = \sqrt{A + B a^{-6}} \,
\end{equation}
where $B$ is an integration constant, and $a = 1/(1+z)$. The total EoS reads:
\begin{eqnarray} \label{omega tot chap}
 \omega_{tot} &=& \frac{ \mathcal{P}}{\rho_m + \rho}
 \nonumber\\&&= - \frac{A}{\rho} \left( \rho + \rho_m \right)^{-1} \nonumber\\&&
 = - \frac{A}{\sqrt{A + B(1+z)^6}} \left[ \Omega_b (1+z)^3 + \sqrt{A + B (1+z)^6} \right]^{-1} \nonumber\\&&
 = - \frac{A}{\Omega_b (1 + z)^3 \sqrt{A + B (1+z)^6} + A + B(1+z)^6} \,
\end{eqnarray}
where $\Omega_b$ here is the baryon matter density, normalized in units of the critical density. Analogously, $\omega_{DE}$ can be written as:
\begin{equation}
 \omega_{DE} = \frac{\mathcal{P}}{\rho} = - \frac{A}{\rho^2} = - \frac{A}{A + B (1 + z)^6}\,
\end{equation}

Finally, we write down the normalized Hubble parameter as:
\begin{equation}
 \mathcal{E}(z) = \sqrt{ \Omega_b (1+z)^3 + \Omega_{DE} \cdot \sqrt{ A+(1-A)(1+z)^6} } \,
\end{equation}

\noindent where $\Omega_{DE}=1-\Omega_b$ is the DE density in critical DE density units. For the Chaplygin gas, the total matter density is not given by the sum of baryons and CDM. In fact, the Chaplygin gas is known as the first example of a unified dark energy model, in which CDM is generated from assuming baryons only within the energy momentum tensor (For more details, see \cite{k-essence-ecc1.1, k-essence-ecc1.2, k-essence-ecc1.3, k-essence-ecc1.4, k-essence-ecc2.1, k-essence-ecc2.2, k-essence-ecc2.3, k-essence-ecc3.1, k-essence-ecc3.2, k-essence-ecc3.3, k-essence-ecc3.4, k-essence-ecc4.1, k-essence-ecc4.2, k-essence-ecc4.3, k-essence-ecc4.4, k-essence-ecc5.1, k-essence-ecc5.2, k-essence-ecc5.3, k-essence-ecc6.1, k-essence-ecc6.2, k-essence-ecc6.3}). The cosmographic parameters read:
\begin{equation}\label{hdysfannn}
q_0 = \frac{1}{2} [1 + 3 A (\Omega_b - 1)]\,
\end{equation}
\begin{equation}\label{jsyayiudygggggg}
j_0= \frac{1}{2} [2+ 9 A (A - 1) (\Omega_b -1 ) ]\,
\end{equation}
and:
\begin{equation}\label{jsyayiudygggggg2}
s_0= \frac{1}{4} \{ -14 -9 A ( \Omega_b -1 ) \{7 + A [3 A (5+\Omega_b ) -3 \Omega_b -20 ]\} \}\,
\end{equation}

Inverting, we have:
\begin{equation}
 \Omega_{m,q_0,j_0} = 2 \cdot \frac{j_0 + q_0 + 2 q_0^2 -2}{2 j_0 + 6 q_0 -5}\,
\end{equation}

\section{The Dvali--Gabadadze--Porrati Model} \label{sec modello DGP}

Here, we discuss the DGP model, which attempts to give an
alternative to a positive cosmological constant, due to a
modification of gravity at large distances \cite{DGP2}. The scale
length is regularized by the typical crossover radius, $r_c$. Below
it, the Einstein gravity is preserved; over it, the gravitational
force follows a five-dimensional $1/r^{3}$ behavior. The DGP model represents an infrared modification of standard gravity. The approach is inspired by
Braneworld constructions with infinite-volume from extra dimensions \cite{DGP2bis}. In other words, these extra dimensions are infinite or, more likely, cosmologically large. It follows that the graviton mass cannot be massless as in standard GR, providing, instead, massive states, whose characteristic width is $r_{c}^{-1}$.

In particular, by assuming to reduce our description to a four-dimensional space-time, we have:
\begin{equation}\label{uno}
3M_{Pl}^2\left(\mathcal{H}^2-{\mathcal{H}\over r_c}\right)=\rho_m(1+z)^3\,
\end{equation}
where the crossover scale is defined as:
\begin{equation}\label{kj98}
r_c\equiv\frac{1}{\mathcal{H}_0(1-\Omega_m)}\,
\end{equation}
Considering the normalized Hubble rate, \emph {i.e.},
$\mathcal{E}(z)\equiv\frac{\mathcal{H}(z)}{\mathcal{H}_0}$, we infer:
\begin{equation}\label{star}
\mathcal{E}(z)=\sqrt{\Omega_{m}(z)+\Omega_{r_c}}+\sqrt{\Omega_{r_c}}\,
\end{equation}
where:
\begin{equation}\label{45}
\Omega_{r_c}=\frac{1}{4r_c^2\mathcal{H}_0^2}\,
\end{equation}
may be interpreted as the density of the crossover factor. It is clear that:
\begin{equation}\label{constraint}
\Omega_{r_c}=\frac{1}{4}\left(1-2\Omega_{m}+\Omega_{m}^{2}\right)\,
\end{equation}
Thus, the model depends on one parameter only. We get:
\begin{eqnarray}\label{fattori}
q_0&=&\frac{1}{2}+\frac{3}{2}\Big[\frac{\Omega_m-1}{\Omega_m+1}\Big]\,\nonumber\\
\,\nonumber\\
j_0&=&\frac{1+\Omega_{m}\left(3-6\Omega_{m}+10\Omega_{m}^{2}\right)}{\left(1+\Omega_{m}\right)^3}\\
\,\nonumber\\
s_0&=&\frac{\left(1-5\Omega_m\right)\left\{1 + \Omega_m \Big\{ 1 + 2
\Omega_m \left[ 12 - \Omega_m \left(7 - 8
\Omega_m\right) \right] \Big\} \right\}}{\left(1+\Omega_m\right)^5}\nonumber\,
\end{eqnarray}
and since the model is one-parameter-dependent, we have:
\begin{equation}
 \Omega_{m,q_0} = \frac{ 1 + q_0 }{2- q_0}
\end{equation}

\subsection{An Extension of DGP: The $\alpha$ Dark Energy Model}
 \label{sec modello aDGP}

A possible phenomenological extension of DGP is the $\alpha$ dark energy model,
hereafter $\alpha$DE \cite{alphaDE}. Since the Braneworld theory, at
large scales, provides that gravity leaks out into the bulk, as a
result, an accelerated expansion of the universe may be related to
them by starting from a modified Friedmann equation. Moreover, phenomenological motivations can extend Equations (\ref{uno}) and (\ref{star}). In particular, it has been proposed in \cite{alphaDE} as an extension of the DGP model. In
this context, the Friedmann equations are modified by using a further parameter, \emph {i.e.}, $\alpha$. We write down:
\begin{equation}
3M_{Pl}^2\left(\mathcal{H}^2-{\mathcal{H}^\alpha\over
r_c^{2-\alpha}}\right)=\rho_m(1+z)^3\,
\end{equation}
and we define:
\begin{equation}\label{errec}
r_c=\frac{(1-\Omega_m)^{\frac{1}{\alpha-2}}}{\mathcal{H}_0}\,
\end{equation}
whereas an implicit expression of $\mathcal{E}(z)$ is:
\begin{equation}\label{blut}
\mathcal{E}(z)=\sqrt{\Omega_m(z)+\mathcal{E}(z)^\alpha(1-\Omega_m)}\,
\end{equation}
It is clear that instead of Equation (\ref{star}), this model accounts for two parameters, \emph {i.e.}, $\Omega_m$ and $\alpha$. The limiting cases
$\alpha=1$ and $\alpha=0$ correspond to the DGP and
$\Lambda$CDM cases, respectively. Even though the physical meaning of $\alpha$ is only phenomenological, it leads to the existence of some hidden and unknown quantum processes. Thus, its introduction relies on the lack of knowledge of quantum processes, occurring at early epochs of the universe's evolution.

The CS for the $\alpha$DE model are:

\begin{eqnarray}\label{ks}
q_0&=&\frac{\Omega_m(3-\alpha)+\alpha-2}{2-\alpha(1-\Omega_m)}\,\nonumber\\
\nonumber\\
j_0&=&\frac{8 + \alpha\left(\Omega_m-1\right)\left(12 + (-6 +
\alpha) \alpha - 2 \left(-3 + \alpha\right) \alpha \Omega_m+\left(-6
+ \alpha\right) \left(-3 +
\alpha\right)\Omega_{m}^{2}\right)}{2-\alpha(1-\Omega_m)}\,\\
\nonumber\\
s_0&=&\frac{-(\alpha-2)^5+(\alpha-2)^4A\Omega_m-(\alpha-2)^3B\Omega_{m}^{2}+2(\alpha-2)^2C\Omega_{m}^{3}-\alpha(\alpha-2)D\Omega_{m}^{4}+\alpha^2 G\Omega_{m}^{5}}{2-\alpha(1-\Omega_m)}\,\nonumber
\end{eqnarray}
with:
\begin{eqnarray}
 A&=& 5\alpha-27\, \nonumber\\
 B&=&63+2\alpha(5\alpha-27) \nonumber\\
 C&=&-27+\alpha[-27+\alpha(-27+5\alpha)]\,\\
 D&=&162+\alpha[-27+\alpha(-54+5\alpha)]\, \nonumber\\
 E&=&-108+\alpha[90+\alpha(27+\alpha)]\,\nonumber
\end{eqnarray}
while the explicit forms of $\Omega_m$ as a function of the CS cannot be inferred from analytical expressions. We will numerically solve the corresponding values of $\Omega_m$ in terms of the cosmographic coefficients, as we report in Table 1.

\begin{table}[ht!]\centering
\caption{{\small Table of cosmographic results. We reported the values of the cosmographic series (CS) by considering four different cosmological fits through the use of supernovae Ia (SNeIa), baryonic acoustic oscillation (BAO) and cosmic microwave background (CMB) measurements. In so doing, we minimized the chi-squared given by Equation (\ref{soam}), and we performed standard Bayesian analyses. In particular, the first fit leaves the coefficients free to vary; the second one assumes $\mathcal{H}_0$ fixed to the PLANCK    
 value. The third fit uses $\mathcal{H}_0$ fixed in terms of the Hubble Space Telescope (HST) measurement, whereas the last fit fixes the value of the Hubble rate today, through the fit of a first order $d_L$ within the interval $z<0.36$ by using SNeIa measurements only. The values of $\Omega_m$, $f_{0}$, $f_{z0}$, $f_{2z0}$, $f(\mathcal{T}_0)$, $f^{''}(\mathcal{T}_{0})$, $f^{'''}(\mathcal{T}_{0})$ and $f^{(iv)}(\mathcal{T}_0)$ have been obtained by using standard error propagations. The value of $f^{'''}(\mathcal{T}_{0})$ for fit 3 is in powers of $10^{3}$, while the value of $f^{(iv)}(\mathcal{T}_0)$ is in powers of $10^8$. Moreover, fit 3 cannot constrain the Chevallier--Polarsky--Linder (CPL) model, showing that this model is disadvantageous for high values of $s_0$.}}

\begin{tabular}{@{}ccccc@{}}
\toprule
\quad \textbf{Parameter} \quad & \quad \textbf{Fit 1} \quad & \quad \textbf{Fit 2} \quad & \quad \textbf{Fit 3} \quad & \quad \textbf{Fit 4} \quad \\
\midrule
 {\small $\chi^2$} & {\small $0.973$} & {\small $1.042$} & {\small $1.040$} & {\small $0.974$} \\ [0.8ex]

 {\small $\mathcal{H}_0$} & {\small $70.087$}{\tiny${}_{-1.266}^{+1.289}$} & {\small $68.000$}{\tiny${}_{-1.400}^{+1.400}$} & {\small $72.000$}{\tiny${}^{+8.000}_{-8.000}$} & {\small $69.960$}{\tiny${}^{+1.120}_{-1.160}$} \\

 {\small $q_0$} & {\small $-0.631$}{\tiny${}^{+0.154}_{-0.123}$} & {\small $-0.561$}{\tiny${}^{+0.137}_{-0.135}$} & {\small $-0.986$}{\tiny${}^{+0.099}_{-0.094}$} & {\small $-0.596$}{\tiny${}^{+0.150}_{-0.120}$} \\

 {\small $j_0$} & {\small $1.571$ }{\tiny${}^{+0.431}_{-0.404}$} & {\small $2.477$}{\tiny${}^{+1.460}_{-1.390}$} & {\small $4.594$}{\tiny${}^{+1.370}_{-1.180}$} & {\small $1.285$}{\tiny${}^{+0.459}_{-0.431}$} \\
 {\small $s_0$} & {\small $3.001$}{\tiny${}_{-2.086}^{+2.279}$} & {\small $6.170$}{\tiny${}^{+6.189}_{-6.157}$} & {\small $23.772$}{\tiny${}^{+5.076}_{-5.048}$} & {\small $1.807$}{\tiny${}^{+2.271}_{-2.080}$} \\
 {\small $\Omega_{m,\Lambda \mathrm{CDM}}$} & {\small $0.246$}{\tiny${}_{-0.082}^{+0.103}$} & {\small $0.293$}{\tiny${}^{+0.091}_{-0.090}$} & {\small $0.009$}{\tiny${}^{+0.066}_{-0.062}$} & {\small $0.270$}{\tiny${}^{+0.100}_{-0.080}$} \\

 {\small $\Omega_{m,\omega \mathrm{CDM}}$} & {\small $0.355$}{\tiny${}_{-0.146}^{+0.171}$} & {\small $0.517$}{\tiny${}_{-0.225}^{+0.234}$} & {\small $0.452$}{\tiny${}^{+0.146}_{-0.131}$} & {\small $0.328$}{\tiny${}^{+0.186}_{-0.161}$} \\
 {\small $\Omega_{m,\mathrm{CPL}}$} & {\small $0.193$}{\tiny${}_{-0.567}^{+0.615}$} & {\small $0.374$}{\tiny${}_{-1.807}^{+1.862}$} & {\small $---$} & {\small $0.207$}{\tiny${}^{+0.504}_{-0.462}$} \\
 {\small $\Omega_{m,\mathrm{Chapl.}}$} & {\small $0.094$}{\tiny${}_{-0.209}^{+0.237}$} & {\small $-0.320$}{\tiny${}_{-1.098}^{+1.153}$} & {\small $-4.112$}{\tiny${}^{+9.181}_{-8.003}$} & {\small $0.200$}{\tiny${}^{+0.222}_{-0.194}$} \\
 {\small $\Omega_{m,\mathrm{DGP}}$} & {\small $0.141$}{\tiny${}_{-0.053}^{+0.067}$} & {\small $0.171$}{\tiny${}_{-0.062}^{+0.063}$} & {\small $0.005$}{\tiny${}^{+0.033}_{-0.032}$} & {\small $0.156$}{\tiny${}^{+0.0670}_{-0.0536}$} \\
 {\small $\Omega_{m,\alpha \mathrm{DGP}}$} & {\small $0.241$}{\tiny${}^{+0.102}_{-0.081}$} & {\small $0.287$}{\tiny${}^{+0.091}_{-0.089}$} & {\small $0.009$}{\tiny${}^{+0.065}_{-0.061}$} & {\small $0.264$}{\tiny${}^{+0.099}_{-0.079}$} \\
 {\small $f_0$} & {\small $-2.584$}{\tiny${}^{+0.247}_{-0.214}$} & {\small $-2.368$}{\tiny${}^{+0.224}_{-0.222}$} & {\small $-3.096$}{\tiny${}^{+0.791}_{-0.785}$} & {\small $-2.541$}{\tiny${}^{+0.229}_{-0.202}$} \\
 {\small $f_{z0} $} & {\small $-3.124$}{\tiny${}^{+1.839}_{-1.666}$} & {\small $-0.232$}{\tiny${}^{+4.440}_{-4.241}$} & {\small $5.002$}{\tiny${}^{+5.681}_{-5.074}$} & {\small $-3.848$}{\tiny${}^{+1.913}_{-1.746}$} \\
 {\small $f_{2z0} $} & {\small $-13.646$}{\tiny${}^{+11.489}_{-10.295}$} & {\small $-26.331$}{\tiny${}^{+26.546}_{-26.142}$} & {\small $-82.383$}{\tiny${}^{+41.063}_{-40.243}$} & {\small $-9.483$}{\tiny${}^{+11.200}_{-10.066}$} \\
 {\small $f(\mathcal{T}_0)$} & {\small $-4.969$}{\tiny${}^{+0.242}_{-0.239}$} & {\small $-4.678$}{\tiny${}^{+0.248}_{-0.248}$} & {\small $-5.244$}{\tiny${}^{+1.228}_{-1.228}$} & {\small $-4.951$}{\tiny${}^{+0.217}_{-0.223}$} \\
 {\small $f''(\mathcal{T}_0)$} & {\small $-0.047$}{\tiny${}^{+0.106}_{-0.088}$} & {\small $-0.013$}{\tiny${}^{+0.073}_{-0.072}$} & {\small $-5.247$}{\tiny${}^{+39.754}_{-37.822}$} & {\small $-0.028$}{\tiny${}^{+0.087}_{-0.073}$} \\
 {\small $f'''(\mathcal{T}_0)$} & {\small $-0.177$}{\tiny${}^{+0.384}_{-0.328}$} & {\small $-0.274$}{\tiny${}^{+0.593}_{-0.576}$} & {\small $-15.943$}{\tiny${}^{+352.379}_{-334.455}$} & {\small $-0.073$}{\tiny${}^{+0.213}_{-0.187}$} \\
 {\small $f^{(iv)}(\mathcal{T}_0)$} & {\small $-1.385$}{\tiny${}^{+3.481}_{-2.979}$} & {\small $-2.608$}{\tiny${}^{+7.240}_{-7.045}$} & {\small $-1.431 $}{\tiny${}^{+52.560}_{-49.860}$} & {\small $-0.573$}{\tiny${}^{+1.354}_{-1.190}$} \\ \bottomrule
\end{tabular}
\vspace{12pt}
\label{tab:fits}
\end{table}

\section{Dark Energy from Extended Theories of Gravity} \label{sec gravita estesa}

In the previous Sections, we treated the case of DE in terms of evolving fluids within the energy-momentum tensor of Einstein's gravity. Here, we consider an alternative point of view based on extensions of GR. In so doing, the DE fluid naturally arises as further corrections from the Hilbert--Einstein action. In particular, alternative theories of standard Einstein's gravity have been first developed to alleviate the problems of standard cosmology, in which, by simply introducing a vacuum energy cosmological constant, it is not possible to characterize the universe dynamics from prime principles \cite{adj}. This class of theories, on the contrary, constitutes a plausible attempt to determine a semi-classical scheme, in which one deems GR as a limiting case of a more general paradigm, described by extensions of the Einstein--Hilbert action.
One of the most commonly used approaches is that of Extended Theories of
Gravity \cite{review.1, review.2, review.3, review.4, review.5}. In these theories, Einstein's Lagrangian is extended and corrected through higher-order curvature invariants
and minimally- or non-minimally-coupled scalar fields (e.g., $\mathcal{R}^{2}$, $\mathcal{R}^{\mu\nu} \mathcal{R}_{\mu\nu}$,
{$\mathcal{R}^{\mu\nu\alpha\beta}\mathcal{R}_{\mu\nu\alpha\beta}$}, $\mathcal{R} \,\Box \mathcal{R}$, or $\mathcal{R}\,\Box^{k}\mathcal{R}$, and $\phi^{2}R$).
Other interesting motivations to extend general relativity have been carried out for understanding the role played by the Mach principle in the framework of a theory of gravity, leading one to consider a varying gravitational coupling. \mbox{A well-known} example is represented by Brans--Dicke theory \cite{BD}, which includes a variable gravitational coupling. The corresponding dynamics is influenced by a single scalar field, non-minimally coupled to the geometry, and so, the above-mentioned Mach's principle is recovered \cite{cimento,odintsov,DMf(R)}.

 In addition, one may imagine extending GR, in comparison with every unification scheme of fundamental interactions \cite{sciama}. Similar approaches have been developed in quantum field theory on curved spacetime, giving interactions between quantum scalar fields and background geometry, or gravitational self-interactions, that actually provide this kind of correction to the Einstein--Hilbert Lagrangian \cite{birrell,vilkovisky}. For the sake of clarity, a self-consistent quantum gravity scheme is, however, not derived from these approaches, which do not constitute a full quantum gravity framework. One of the simplest extensions to GR is $f(\mathcal{R})$
gravity in which the Lagrangian density is an arbitrary function of Ricci scalar $\mathcal{R}$~\cite{defelice}. The first example, Starobinsky's model for inflation, $f(\mathcal{R})=\mathcal{R}+\alpha \mathcal{R}^2$ ($\alpha>0$), could lead to the accelerated expansion of the universe due to the presence of $\alpha \mathcal{R}^2$ \mbox{term \cite{Star80}}. Furthermore, this model's results are consistent with the
temperature anisotropies observed in CMB, and then, it seems to be a viable
alternative to the scalar-field models of inflation \cite{defelice}. $f(\mathcal{R})$-models to describe DE behavior are studied by \cite{rev5, fRearly2p,fRearly2, fRearly3, fRearly4,Dolgov,OlmoPRL, Olmo05, Erick06, Chiba07, Navarro, CapoTsuji}. Furthermore, models that satisfy both Solar System and cosmological constraints have been proposed in~\cite{AGPT, LiBarrow, AmenTsuji07, Hu07, Star07, Appleby,Tsuji08, Natalie, Cognola07, Linder09,ivan,fabio,quadru,odi,odi1.1, odi1.2, odi1.3}. In particular, since the Newtonian law on large scales undergoes changes in $f(\mathcal{R})$-gravity, this leaves several interesting observational signatures, such as the modification of rotational curves of a galaxy, galaxy clustering spectra \cite{matterper1,matterper2,SongHu1,Pogosian,annalen.1, annalen.2}, CMB~\cite{Zhang05,LiBarrow} \mbox{and weak} lensing~\cite{TsujiTate, Schmidt}. Another alternative approach is based on teleparallel gravity \cite{pereira}. This theory corresponds to a gauge theory of a translation group based on Weitzenb$\ddot{o}$k geometry. In teleparallel gravity, gravitation is attributed to torsion, which plays the role of force~\cite{59}. In teleparallel gravity, this means that no geodesics exist, and force equations are analogous to the Lorentz force equations of electrodynamics. Thus, those force equations describe the particle trajectories submitted to a gauge gravitational field. This means that the gravitational interaction can be described alternatively in terms of curvature, as usually performed in GR, or in terms of torsion, as in teleparallel gravity. One of the first models in $f(\mathcal{T})$-gravity was proposed to avoid the Big Bang singularity and recover inflation without the need for an inflaton \cite{62}.
Several cosmological applications in $f(\mathcal{T})$-gravity to describe the accelerated expansion of the universe are studied in \cite{64,65,66,68,70,71}.

 In what follows, we limit our attention to the case of $f(\mathcal{R})$ and $f(\mathcal{T})$ gravities, and we discuss their roles in terms of cosmography.

\newpage
\subsection{Cosmography of $f(\mathcal{R})$ Gravity} \label{sec f(R)}

In this subsection, we briefly show how to relate $f(\mathcal{R})$ gravity to cosmography. We start from assuming that the Hilbert--Einstein action is replaced by:
\begin{equation}\label{lkjn}
{\cal{A}} = \int{d^4x \sqrt{-g} \left[ f(\mathcal{R}) + {\cal{L}}_{m} \right] }\,
\end{equation}
where ${\cal{L}}_{m}$ is the standard matter term and $g$ is the determinant of the metric tensor. By varying Equation~(\ref{lkjn}) in terms of $g_{\mu\nu}$, we get the following field equations \cite{review.1, review.2, review.3, review.4, review.5}:
\begin{equation}\label{filedeqs}
\mathcal{R}_{\mu \nu}f' (\mathcal{R})-\frac{1}{2}f(\mathcal{R})g_{\mu
\nu}-(\nabla_{\mu}\nabla_{\nu}-g_{\mu
\nu}\nabla_{\alpha}\nabla^{\alpha})f' (\mathcal{R}) =8\pi T_{\mu \nu}\,
\end{equation}
having a curvature DE fluid, associated with $f(\mathcal{R})$ gravities. Here, the prime indicates the derivative of $f(\mathcal{R})$ with respect to the Ricci scalar and $T_{\mu \nu}$ is the energy momentum tensor of matter. Particularly, one can assume that the $f(\mathcal{R})$ class should reduce to $\Lambda$CDM at $z\ll1$, to guarantee the validity of cosmography.
By using the FRW metric within Equation (\ref{filedeqs}), we get the modified Friedmann equations:
\begin{equation}
\mathcal{H}^2 = \frac{1}{3} \left [ \rho_{curv} + \frac{\rho_m}{f'(\mathcal{R})} \right
]\, \label{eq1}
\end{equation}
where $ \rho_{curv}$ is an effective curvature fluid and:
\begin{equation}\label{eq2}
2 \dot{\mathcal{H}} + 3\mathcal{H}^2= - \mathcal{P}_{curv}\,
\end{equation}
giving a DE term, expressed in terms of curvature, \emph {i.e.}:
\begin{equation}\label{eq:rhocurv}
\rho_{curv} = \frac{1}{f'(\mathcal{R})} \left \{ \frac{1}{2} \bigg[ f(\mathcal{R}) - \mathcal{R}
f'(\mathcal{R}) \bigg] - 3 \mathcal{H} \dot{\mathcal{R}} f''(\mathcal{R}) \right \} \,
\end{equation}
with the curvature barotropic factor:
\begin{equation}
\omega_{curv} = -1 + \frac{\ddot{\mathcal{R}} f''(\mathcal{R}) + \dot{\mathcal{R}} \left [ \dot{\mathcal{R}}
f'''(\mathcal{R}) - \mathcal{H} f''(\mathcal{R}) \right ]} {\left [ f(\mathcal{R}) - \mathcal{R} f'(\mathcal{R}) \right ]/2 - 3
\mathcal{H} \dot{\mathcal{R}} f''(\mathcal{R})}\, \label{eq: wcurv}
\end{equation}
and pressure:
\begin{equation}
\mathcal P_{curv}=\frac{1}{2}\left(\mathcal{R}-\frac{f}{f'} \right)+\frac{1}{f'} \left[ f'' \left(2 {\cal H} {\dot {\cal R}}+{\ddot {\cal R}}\right)+{\dot {\cal R}^2f'''}\right]\,
\end{equation}

By means of the above equations, it is clear that the DE problem could be, in principle, addressed by curvature. Further, once $f(\mathcal R)$ and derivatives are bounded by cosmography, the corresponding DE term is naturally constrained, and one can describe the DE effects in terms of observable quantities. Hence, in order to relate $f(\mathcal R)$ and its derivatives to cosmography, we can write down $\mathcal R$ and its derivatives in terms of the redshift, $z$, as:
\begin{equation}\label{Rinz0}
 \begin{split}
 \frac{\mathcal{R}_0}{6} =\, & \mathcal{H}_0\left[\mathcal{H}_{z0}-2\mathcal{H}_0\right]\,\\
 \frac{\mathcal{R}_{z0}}{6} =\,& \mathcal{H}_{z0}^2+\mathcal{H}_0(-3\mathcal{H}_{z0}+\mathcal{H}_{2z0})\,\\
 \frac{\mathcal{R}_{2z0}}{6} =\,& -2\mathcal{H}_{z0}^2+3\mathcal{H}_{z0}\mathcal{H}_{2z0}+\mathcal{H}_0(-2\mathcal{H}_{2z0}+\mathcal{H}_{3z0})\,
\end{split}
\end{equation}
Thus, since $\mathcal R=\mathcal R(z)$, we obtain:
\begin{equation}\label{f0fz0fzz0dopo}
\begin{split}
\frac{f_0}{2\mathcal{H}_0^2}=\,&-2+q_0\,\\
\frac{f_{z0}}{6\mathcal{H}_0^2} =\,&-2-q_0+j_0\,\\
\frac{f_{2z0}}{6\mathcal{H}_0^2}=\,&-2-4q_0-(2+q_0)j_0-s_0\,
\end{split}
\end{equation}
which represent the cosmographic constraints on $f(z)=f(\mathcal R(z))$.

%
%
\subsection{Cosmography of $f(\mathcal{T})$ Gravity} \label{sec f(T)}

As we described above, an alternative view can be achieved by considering torsion. In particular, models based on $f(\mathcal T)$ gravity include torsion as the source of DE \cite{pereira, storny.1, storny.2, storny.3, storny.4, bamba}. The general action~is:
\begin{equation}
{\mathcal A}=
\int d^4x \abs{e} \left[ \frac{f(\mathcal T)}{2{\kappa}^2}
+{\mathcal{L}}_{\mathrm{M}} \right]\,
\label{eq:IXA-2.6}
\end{equation}
where $e_A (x^{\mu})$ is an orthonormal basis for the tangent space at each point, $x^{\mu}$, on a generic manifold, with $\mathcal{A}$ running over zero, one, two and three. Moreover, the metric is given by $
g_{\mu\nu}=\eta_{A B} e^A_\mu e^B_\nu$, where $\mu$ and $\nu$ are the coordinate indices on the manifold. The torsion and co-torsion, respectively, read:
\begin{equation}
\mathcal T^\rho_{\verb| |\mu\nu} \equiv e^\rho_A
\left( \partial_\mu e^A_\nu - \partial_\nu e^A_\mu \right)\,
\label{eq:IXA-2.2}
\end{equation}
and:
\begin{equation}
K^{\mu\nu}_{\verb| |\rho}
\equiv
-\frac{1}{2}
\left(\mathcal T^{\mu\nu}_{\verb| |\rho} - \mathcal T^{\nu \mu}_{\verb| |\rho} -
\mathcal T_\rho^{\verb| |\mu\nu}\right)\,
\label{eq:IXA-2.3}
\end{equation}
while one infers the torsion scalar, $\mathcal T \equiv S_\rho^{\verb| |\mu\nu} \mathcal T^\rho_{\verb| |\mu\nu}$, where:
\begin{equation}S_\rho^{\verb| |\mu\nu} \equiv \frac{1}{2}
\left(K^{\mu\nu}_{\verb| |\rho}+\delta^\mu_\rho \
\mathcal T^{\alpha \nu}_{\verb| |\alpha}-\delta^\nu_\rho \
\mathcal T^{\alpha \mu}_{\verb| |\alpha}\right)\,
\end{equation}
is the superpotential.

The corresponding field equations read:
\begin{equation}\label{eq:IXA-2.7}
\frac{{\kappa}^2}{2} e_A^\rho
{\mathcal T^{(\mathrm{M})}}_\rho^{\verb| |\nu} = \frac{1}{e} \partial_\mu \left( eS_A^{\verb| |\mu\nu} \right) f^{\prime}-e_A^\lambda \mathcal T^\rho_{\verb| |\mu \lambda} S_\rho^{\verb| |\nu\mu}
f^{\prime}
 + S_A^{\verb| |\mu\nu} \partial_\mu \left(\mathcal T\right) f^{\prime\prime}
+\frac{1}{4} e_A^\nu f \,\nonumber
\end{equation}
\noindent where ${\mathcal T^{(\mathrm{M})}}_\rho^{\verb| |\nu}$ represents the energy-momentum tensor. The modified Friedmann equations can be rewritten as:
\begin{eqnarray}
\mathcal{H}^2& =& \frac{1}{3} \left[ \rho+ \rho_{\mathcal T} \right]\, \label{eq1} \\
2 \dot{\mathcal{H}} + 3\mathcal{H}^2&=& - \frac{1}{3} \left( \mathcal{P}+\mathcal{P}_{\mathcal T} \right) \,\label{eq2}
\end{eqnarray}
with the DE density given by:
\begin{equation}\label{rhoT}
 \rho_{\mathcal T}= \frac{1}{2} \left[ 2{\mathcal T}f'({\mathcal T})-f({\mathcal T})-{\mathcal T}/2 \right] \,
\end{equation}
and the barotropic factor:
\begin{equation}\label{omegaeff}
 \omega_{ \mathcal{T}} \equiv \frac{ \mathcal{P}_{\mathcal{T}} }{ \rho_{\mathcal{T}} } = -1 + \frac{ 4\dot{\mathcal{H}} (4 \mathcal{T} f''( \mathcal{T} ) + 2f' (\mathcal{T})-1)}{4 \mathcal{T}f'( \mathcal{T} )- 2 f( \mathcal{T})- \mathcal{T}} \,
\end{equation}

This result could be, in principle, related to the observed acceleration of the universe.
Finally, we~find,
\begin{eqnarray}\label{fT}
 f( \mathcal{T}_0) &=& 6 \mathcal{H}_0^2 \,(\Omega_{m}-2)\, \nonumber
 \end{eqnarray}
 \begin{eqnarray}
 f''( \mathcal{T}_0) &=& \frac{1}{6 \mathcal{H}_0^2} \left[\frac{1}{2}-\frac{3\Omega_{m}}{4(1+q_0)}\right] \, \nonumber\\
 \end{eqnarray}
 \begin{eqnarray}
 f'''(\mathcal{T}_0) &=& \frac{1}{(6\mathcal{H}_0^2)^{2}}\left[\frac{3}{4}-\frac{3\Omega_{m}(3q_0^2+6q_0+j_0+2)}{8(1+q_0)^3}\right]\,\, \nonumber
 \end{eqnarray}
 \begin{eqnarray}
 f^{(iv)}(\mathcal T_0) &=& -\frac{1}{(6\mathcal{H}_0^2)^{3}}\left\{ \frac{3\Omega_{m}}{16(1+q_0)^5}\,\left[s_0(1+q_0)+j_0(6q_0^2 +17q_0+3j_0+5) +\right.\right.\nonumber\\&& \left.\left. +3q_0(5q_0^3+20q_0^2+29q_0+16)+9 \right]+\frac{15}{8} \right\}\,\nonumber
\end{eqnarray}
and $f'(\mathcal{T}_0)=1$ to guarantee Solar System constraints.

Summing up, the case of extended theories of gravity is at most different from usual DE models. Here, the role played by the introduction of DE is replaced by additional terms proportional to curvature, torsion or quantum effects, and so forth. However, as we pointed out in previous sections, instead of considering the matter density, we focus our attention on constraining the additional terms that extend the Einstein--Hilbert action. In the next sections, we will discuss the role of experimental procedures in cosmology, and we highlight the possibility of relating those techniques to cosmographic analyses. \mbox{We therefore} get constraints on the observable universe, deriving the cosmographic coefficients and the matter densities for each model in terms of the CS. As concern $f(\mathcal R)$ and $f(\mathcal T)$ gravities, we will constrain their values and the corresponding derivatives in terms of CS.

%
\section{Experimental Procedures} \label{sec procedure sperimentali}

Now, we want to emphasize the role played by three of the most relevant standard candles in cosmology. Once we give information on the experimental techniques based on supernovae, CMB and BAO, we relate those methods to the cosmographic analysis, and we infer present numerical values by directly minimizing the corresponding chi-squared. Let us first start by noticing that one of the main challenges of modern cosmology is to reconstruct the behavior of $a(t)$, for different eras of the universe's evolution. Indeed, reproducing the universe expansion history at different epochs may represent a way of discriminating among models which one behaves better than others. If we limit our attention to present times, another interesting point is even to find out at which redshift the passage, between that matter- to the DE-dominated epochs, occurred.

The corresponding redshift is known in the literature as \textit{transition redshift}, as we previously stressed. The transition redshift and the corresponding observational experimental consequences depend on the specific form of DE, and so, alleviating the degeneracy problem may represent a viable procedure to fix tighter constraints on the expansion history of the universe. One of the possibilities, as mentioned in Section \ref{sec intro}, is to perform a superpositions of different cosmological tests, reducing the allowed phase space diagrams and, thus, finding more accurate bounds. Unfortunately, small and high redshift data dominate over intermediate surveys, showing the fact that at redshift $2<z<1,000$, we cannot fix relevant constraints on the expansion history of the universe, with the same precision of small and high redshifts. This problem may be alleviated assuming that other non-conventional standard candles, such as gamma ray bursts, galaxy clusters luminosities, and so forth, may be thought of as non-standard indicators, albeit able to characterize intermediate bounds of the universe's expansion history. Even though this argument is so far the object of intensive studies, a complete and self-consistent way to characterize data from non-conventional standard candles remains a present open question. Our main purpose is to summarize, here and in what follows, three of the major techniques of data fitting, \emph {i.e.}, SNeIa, BAO and CMB. While the first two tests are limited to intervals of $z\leq1.4$, the third test is associated with $z\sim 1100$, allowing for an accurate matching between small and high redshifts.

However, the discussion on possible non-conventional datasets and the understanding of whether they can be used as standard candles are beyond the purposes of this review.
%
%
\subsection{The Observational Problem} \label{sec problemi osservativi}

The use of cosmological distances as indicators relies on assuming the validity of a causal effect between photons emitted and measured. Moreover, to deduce the universe's expansion history, several observational procedures exist. Examples are SNeIa observations, the study of BAO, the large-scale structures, the weak lensing, understanding the abundances and evolution of clusters, the CMB measurements, and so forth \cite{Linder3-4,Linder5,Durrer1}. It is a matter of fact that SNeIa represent standard candles to check the cosmic expansion effects at low redshift, showing the existence of a phase transition between a decelerating and accelerating universe \cite{SNeIa1.1, SNeIa2, Escamilla1.4}. In order to get a direct relation between observational redshift-distance, one gets:
\begin{equation} \label{prima}
 \tau (z) = \int_{a_*}^1 \mathrm{d} a / (a^2 \mathcal{H}) = \int_0^{z_*} \mathrm{d} z / \mathcal{H}(z) \,
\end{equation}
where $\mathrm{d} \tau (z)= \mathrm{d} t / a$ represents the conformal time, and from now on, $a_*=a(z_*)$ is the scale factor at which a photon (event) is emitted. Invoking the validity of the Friedmann equations and the fact that $-1<\omega<0$ (excluding particular cases in which $\omega<-1$ \cite{Escamilla}), the expansion function, $a(t)$, is \cite{Linder}:
\begin{equation}
 a(t) = \int_a^1 \mathrm{d} a' / (a' \mathcal{H}) = \int_0^z \mathrm{d} z' / [(1+z')\mathcal{H}(z')] \,
\end{equation}
and moreover:
\begin{equation}
 \mathcal{H}_0 \tau(z) = \int_0^z \mathrm{d} z' \left\{\Omega_m (1+z')^3 + \Omega_{\mathrm{DE}} \cdot \exp \left[ 3 \int_0^{\ln(1+z')} [1 + \omega(z'')] \mathrm{d} \ln(1+z'') \right] \right\} \,
\end{equation}
By looking at the above equations, it is clear that once a scalar field is determined for a particular model, we could extract the corresponding EoS, although this does not allow one to consider a completely model-independent parameter space.

Taking into account observational data, it is possible to characterize different constraints in which the EoS can be folded. By considering the conformal time, $\tau$, we have $d = (1+z) \tau$, and no foreknowledge and local approximation are required in order to have the relation between conformal time and the scale factor. The error estimation on the observed magnitudes is:
\begin{equation}
 \frac{\sigma_\tau}{\tau} = \sigma_m \frac{\ln 10}{5} \,
\end{equation}
whereas the logarithmic derivative of $a(\tau)$ is:
\begin{equation}
 \frac{\mathrm{d} \tau}{\mathrm{d} \ln a} = \frac{\mathrm{d} t}{ \mathrm{d} a } = (\dot{a})^{-1} \,
\end{equation}
which gives the proper time evolution of the scale factor and the conformal horizon scale, $(a\mathcal{H})^{-1}$. \mbox{The main} advantage of the cosmic time is to relate the redshift in terms of a time variable, giving a possible match between distance and time measurements. In what follows, we are going to describe, in detail, the datasets commonly used in the literature, in terms of the redshift, $z$, and cosmic time, $a(\tau)$.

\subsection{Baryonic Acoustic Oscillations} \label{BAO th}

\textit{Baryonic acoustic oscillations} are oscillations with a wavelength of $ \sim$$0.06h$ Mpc$^{-1}$ (where $h$ is the normalized expansion parameter, $\mathcal{H}= 100 h$), occurring in the power spectrum of matter fluctuations after the recombination epoch. Such a wavelength is related to the comoving sound horizon at the baryon dragging epoch. In the early universe, beyond $z\simeq 1090$, matter and energy have high density values, and hydrogen recombination is not allowed. In the relativistic plasma, large Thomson scattering cross-sections of free electrons and photons occurred in a time lower than the Hubble time. Hence, while radiation pressure tends to separate baryonic matter, gravity tends to cluster it. Oscillations as sound waves in the baryonic-photon fluid were present, due to the fact that radiation pressure is larger than gravitational forces \cite{Peebles and Yu 1970}. Signatures of BAO are present after the recombination. In this phase, the density decreases, and baryonic perturbations can follow the gravitational instability, as well as DM perturbations. Since BAO can be considered sound wave perturbations, it is possible to investigate the corresponding effects through different linear basis sets
\cite{Hu and Sugiyama (1996), Eisenstein and Hu (1998),Meiksin, Seo and Eisenstein 2005, Angulo2005, Springel2005, Jeong, Huff, Angulo2008, Eisenstein et al. (2007b),Weinberg}.

%
%
\subsubsection{Observations of Baryonic Acoustic Oscillations} \label{BAO obs}

Measurements of BAO are due to the spectroscopy of the Sloan Digital Sky Survey (SDSS) data galaxy sample. In particular, SDSS observes different slices in the redshift, constraining the \linebreak distance-redshift relation observed at different epochs. To determine BAO, we introduce the dilatation-scale distance \cite{Seo and Eisenstein 2005, Percival2007c}:
\begin{equation}\label{pippo}
 D_V(z) = \left[ D(z)^2 \frac{z}{\mathcal{H}(z)} \right]^{1/3} \,
\end{equation}
where $D(z) \equiv d_L(z)/(1+z) = \int{\mathrm{d} z/\mathcal{H}(z)}$. When a flat geometry is assumed, we can relate $d_L$ to the angular diameter distance through the following relation: $d_L (z) = (1 + z)^2 D_A(z)$. In particular, the position of the BAO peak approximately constrains $d(z)$, \emph {i.e.}, the ratio of the sound horizon in the decoupling epoch on the dilatation-scale distance:
\begin{equation} \label{ratio}
d(z) = \frac{r_s(z_{ls})}{D_V(z)} \,
\end{equation}
The equation for the comoving sound horizon, $r_s(z_{ls})$, is:
\begin{equation} \label{sound horizon}
 r_s(z_{ls}) = \int_{z_{ls}}^{+ \infty} \frac{c_s(z)}{\mathcal{H}(z)} \mathrm{d} z \,
\end{equation}
where $c_s$ is the sound speed in the early plasma and $z_{\mathrm{ls}}$ the redshift in the dragging epoch (last \mbox{scattering surface).}

Typically, to estimate $D_V (z)$ and $D_A (z)$, the use of cosmic priors is required, indicated by the vector, $\mathbf{Y}$. The parameters depending on the particular cosmological model are indicated by $\mathbf{P}$. \mbox{Therefore, an extended} form for $D_V$ is:
\begin{equation}
 D_V (\mathbf{Y}; \mathbf{P}) = \left[ (1+z)^2 D_A^2 \frac{ z}{\mathcal{H}( \mathbf{Y}; \mathbf{P})} \right]^{1/3} \,
\end{equation}
where, re-writing $D_A$:
\begin{equation}
 D_A (\mathbf{Y}; \mathbf{P}) = \frac{1}{1+z} \int_0^z \frac{\mathrm{d} z'}{\mathcal{H}(\mathbf{Y}; \mathbf{P})} \,
\end{equation}
Percival \textit{et al}. \cite{Percival2009} found an almost independent constraint on the ratio of distances, $D_V(0.35)/D_V(0.2)$, which is consistent at the $1.1 \sigma$ level with the best fit of the $\Lambda$CDM model. A correlation covariance matrix, $\mathbf{C}$, is introduced to define the chi-squared function of BAO:
\begin{equation}
 \chi^2_{\mathrm{BAO}}(\mathbf{P}) = \mathbf{X^T_{\mathrm{BAO}}}(\mathbf{P}) \mathbf{C^{-1}} \mathbf{X_{\mathrm{BAO}}}(\mathbf{P}) \,
\end{equation}
Equation (\ref{pippo}) allows one to connect the Hubble parameter to $D_V$ and $D(z)$:
\begin{equation}\label{maronx}
 \mathcal{H}(z) = \frac{z D(z)^2}{D_V(z)^3} \,
\end{equation}
Equation (\ref{maronx}) permits us to measure $d_L$ and $D_V$ in order to obtain $\mathcal{H}(z)$ \cite{Shafieloo26}. Particularly, \linebreak Shafieloo \textit{et al}. \cite{Shafieloo} showed that the ratio between the Hubble parameter at different epochs plays an important role in cosmological analyses. In fact, one can write down:
\begin{equation}
 \mathcal{H}(z_i,z_j) \equiv \frac{\mathcal{H}(z_i)}{\mathcal{H}(z_j)} = \frac{z_i}{z_j} \left[ \frac{D(z_i)}{D(z_j)} \right]^2 \left[ \frac{D_V(z_j)}{D_V(z_i)} \right]^3 \,
\end{equation}
or equivalently:
\begin{equation} \label{equivalente}
 \mathcal{H}(z_i,z_j) \equiv \left( \frac{z_j}{z_i} \right)^2 \left[ \frac{D(z_i)}{D(z_j)} \right]^2 \left[ \frac{A(z_j)}{A(z_i)} \right]^3 \,
\end{equation}
where $A(z)$ is the BAO acoustic parameter, \emph {i.e.}:
\begin{equation} \label{acoustic}
A(z) = \frac{100 D_V(z) \Omega_{m} h^2}{z}\,
\end{equation}
the ratios among $D_V$, $A$ and $d(z)$ are also related to the ratio of the Hubble parameters at different $z$. The importance of Equation (\ref{equivalente}) is that $\mathcal{H}(z_i ; z_j )$ does not depend either on $\mathcal{H}_0 \sqrt{\Omega_{0m} h^2}$ or on $r_s (z_{ls} )$.
%
\subsection{The Supernova Ia Measurements}

Supernovae type Ia may be considered the most fruitful standard rulers known in the literature, so far. The mechanism behind supernova explosions is complex and particular for each event, and although the luminosity is different for each supernova \cite{Phillips}, the Phillips relation relates the B magnitude peak to the luminous decay after $15$ {days} (however, some exceptions exist, e.g., SN2003fg by Howell \cite{Howell}). Supernovae are distributed at different $z$, crossing two eras of the universe, at which it is possible to distinguish between decelerating and accelerating phases. In fact, the whole interval spans from $z=0$ to $z \lesssim 2$. They are classified as ``Type Ia'', due to the absence of hydrogen and the presence of once ionized silicon (SiII) in their early-time spectra \cite{Filippenko}. The more accredited explanation is that they arise from thermonuclear explosions of white dwarfs, once Chandrasekhar's limit exceeds,\emph {i.e.}, $\sim$$1.4 M_{\odot}$. Even though a self-consistent explanation of the internal process is not completely clarified, supernovae can be found in all galaxies \cite{Barbon}, except in the arms of spiral galaxies \cite{Van Dyk}.


\subsubsection{Supernova Ia Observations}

SNeIa catalogs are typically used and frequently improved. Recent examples are Union 1 \cite{Kowalski}, Union 2 \cite{Amanullah}, the SNLS3
 \cite{Salzano8.1, Salzano8.2, Salzano8.3} compilations, and so on. We rewrite the luminosity distance as:
\begin{equation} \label{dL SN}
 d_L (z, \mathbf{Y}, \mathbf{P}) = \frac{(1+z)}{\mathcal{H}_0} \int_0^z \frac{\mathrm{d} z'}{\mathcal{H}(z', \mathbf{Y}, \mathbf{P})} \,
\end{equation}
The modulus distance reads \cite{Linder,Salzano,Escamilla}:
\begin{equation}
 \mu(z) = 5 \log_{10} (d_L) + \mu_0 \,
\end{equation}
where $\mu_0$ is linked with the absolute magnitude. The error on $z$ is usually negligibly small, while the error on $\mu$, $\delta\mu$ is:
\begin{equation}
 \frac{\Delta d_L}{d_L} = \frac{\log (10)}{ 5} \Delta \mu \,
\end{equation}
The best fit is obtained by minimizing the $\chi^2$ function:
\begin{equation}
 \chi^2 (\mu_0, \mathbf{P}) = \sum_{i=1}^{N_s} \frac{\mu_i(z_i, \mathbf{Y}, \mu_0, \mathbf{P}) - \mu_{obs,i}(z_i)}{\sigma^2_{\mu,i}} \,
\end{equation}
The sums run on the number of the sample objects, $N_s$, while $\mu_i$ represents the index and $i$-th represents the generic supernova. In our convention, $\mu_{obs,i}$ are supernova measurements and $\sigma_{\mu,i}$ measurement~variances.

\subsection{The Cosmic Microwaves Background Measurement}

The CMB radiation was discovered by Penzias and Wilson in 1965 \cite{WilsonPanzias}. The source of this radiation is placed at the time of the last scattering surface. At higher $z$, matter and radiation are mixed together, giving a plasma of photons and electrons with short mean free paths. Radiation is trapped, and matter is highly ionized \cite{COBE Fixsen}. Present temperature is around $T\sim 2.73\,$K, with small fluctuations depending on baryon-photon oscillations, that on comoving scales $\lesssim$$8 h^{-1}$ Mpc are known as Silk damping \cite{Silk}. This leads to an anisotropy of matter density, and the surface of the last scattering consequently emits with different photons, because of the Sachs--Wolfe effect \cite{Sachs-Wolfe}. The Sachs--Wolfe effect is limited to \linebreak large-scale perturbation, and the DE presence dominates in the expansion history, modifying the positions of acoustic peaks in the CMB. For the presence of such density anisotropies, which are translated in terms of temperature fluctuations, it is possible to make a spherical harmonic expansion, due to the fact that the CMB emits from a spherical surface. We have:
\begin{equation}
 \frac{\delta T}{T}(\theta, \phi) = \sum_{l,m} a_{l,m} Y_{l,m}(\theta, \phi) \,
\end{equation}
where $Y_{l,m}(\theta, \phi)$ are Bessel spherical functions, while \mbox{$a_{l,m} = \alpha_{l,m} + i \beta_{l,m}$} depends on the Gaussian statistics. The expectation value of the angular power spectrum is given by \mbox{$C_l = <|a_{l,m}|^2>$}. The corresponding signal is decomposed through the $l$ number, which divides the sphere into the same \mbox{pole numbers.}

If we want to observe a length $L$ on the last scattering surface, we should have: $L=r_0 \theta$, where \mbox{$r_0 = 2c/(\Omega_0 \mathcal{H}_0)$}. If $l$ increases, then $\theta$ gets smaller; so, the latter can contain super galaxy clusters ($ \theta \sim 20'$ with, $l \approx 550$), galaxy clusters ($2' \lesssim \theta \lesssim 10'$, with $l \approx 5500 \div 1200$) or galaxies ($\theta < 1'$, with $l \gtrsim 10000$).

We conclude this section, by underlining that recent CMB data can be either found in several \mbox{works \cite{NewPlanck.1, NewPlanck.2, NewPlanck.3} }or in the \textit{Planck Legacy Archive} \cite{PLA}, which collects all PLANCK's   
 results. In particular, an important result of the PLANCK    
 mission is an anomaly in the observations, leading to the fact that a part of the sky really contains the \textit{Cold Spot} \cite{Vielva-Cruz}. This anomaly was, however, still observed by previous WMAP
 missions \cite{Spergel}, and the thusly obtained PLANCK   
 results corroborate previous discoveries. For the sake of clarity, it was not clear if WMAP measurements were \linebreak appreciable \cite{CriticheWMAP.1, CriticheWMAP.2}, and PLANCK 
 played, therefore, a role of referee in order to understand if this spot really exists or not.
%
\subsubsection{Cosmic Microwaves Background Observations}

The importance of CMB is even correlated to DE constraints, since it is possible to provide relevant constraints at high redshifts. The amplitudes of the acoustic peaks in the CMB angular power spectrum depend sensitively on the locations of the peaks and on matter density. A possible test is performed by using the \textit{CMB shift parameter}, \emph {i.e.}:
\begin{equation}\label{pace}
 R_{\mathrm{CMB}} = (\Omega_b + \Omega_{DM})^{1/2} \int_0^{z_{ls}} \mathcal{H}_0 \frac{\mathrm{d} z'}{\mathcal{H}(z')} \,
\end{equation}
where $\Omega_b$ and $\Omega_{DM}$ are, respectively, baryonic and DM densities and $z_{ls}$ is the redshift at the last \mbox{scattering surface.}

Possible drawbacks would be associated with Equation (\ref{pace}), when one tests unified cosmological models \cite{Orlando62}, since the corresponding Hubble flow can significantly be modified when DE is not negligible for high redshifts. Moreover, in the case of unified models, the sum $\Omega_b+\Omega_{DM}$ in \mbox{Equation (\ref{pace})} reduces to $\Omega_b$, due to fact that DM contributions naturally emerge. The fluid that permits the universe to accelerate at late times is able, in fact, to contribute to the total matter density, at early times. In other words, such a fluid behaves as a cosmological constant at present time and as a pressureless matter term at early times. This is the special case of Chaplygin gas, which we have treated in previous sections. Instead of considering Equation (\ref{pace}) to be consistent with all the cosmological paradigms we illustrated, we replace $R_{\mathrm{CMB}}$ through the ratio $2 l_1/l'_1$. Here, $l_1$ is the position of the first peak on the CMB TT
 power spectrum, while $l'_1$ is the first peak in a Einstein--de Sitter universe with $\Omega_b=1 - \Omega_{DM}$. As a matter of fact, in the $\Lambda$CDM case, the shift parameter naturally reduces to \mbox{Equation (\ref{pace}), }since the cosmological constant contribution is negligible at early times.

We can write $l_1$ as:
\begin{equation}
 l_1 = \frac{D_A(z_{ls})}{r_s(z_{ls})} \,
\end{equation}
where $D_A(z_{ls})$ is known in the literature as \textit{co-moving angular distance}, \emph{i.e.}:
\begin{equation}
 D_A(z_{ls}) = \int_0^{z_{ls}} (1 + z') \mathrm{d} z' \,
\end{equation}
whereas $r_s(z_{ls})$ is the horizon sound, \emph{i.e.}, Equation (\ref{sound horizon}). Moreover, the sound speed reads:
\begin{equation}
 c_s(z) = \left( 3 + 4 \frac{\rho_b}{\rho_\gamma} \right)^{-1/2} \,
\end{equation}
with $\rho_\gamma$ the photon density. It is possible to minimize the following $\chi$-square function:
\begin{equation}
 \chi^2 = \left( \frac{R_{CMB} - R_{CMB,obs}}{\sigma_R} \right)^2 \,
\end{equation}
finding alternative constraints to SNeIa and BAO at higher redshifts. The importance of CMB relies on the fact that a certain model can be checked at the last scattering surface, showing if the DE effects are really relevant or not at early times. In the next section, we find out numerical constraints by using cosmography, combining the three datasets, \emph{i.e.}, SNeIa, BAO and CMB.

\subsection{Cosmographic Fits}

The possibility of performing direct fits in order to infer the CS by means of the SNeIa, BAO and CMB measurements is presented here. In so doing, we highlight the experimental procedures we follow and the role played by error propagations.
The underlying philosophy of performing the cosmographic analysis through the use of the aforementioned cosmological data is intimately related to the idea of building up a sort of \emph{cosmographic test}.

First, let us notice that matching cosmography with observables could be performed by ordering the CS in a hierarchical way. This prescription leads to the choice of fixing parameters and determining a maximum order among them. We therefore consider the sets:
\begin{eqnarray}\label{paurzx}
 \text{A} = \{\mathcal{H}_0, q_0, j_{0},s_0\} \,, \qquad\text{B} =\{ q_0, j_{0},s_0\} \,
\end{eqnarray}
and we fit the CS in terms of the above cited data, by using the luminosity distance, $d_L$, expressed in terms of the redshift, $z$. In this review, we used only the redshift, $z$, instead of parametric variables $y_1$, $y_2$, $y_3$ and $y_4$. We performed the analysis through the use of $z$ only, for brevity. A more complete and accurate analysis can be found in \cite{orlandoluongo}.

Our fits have been carried out by combining the three datasets, SNeIa, BAO and CMB, and minimizing on a grid the corresponding total chi-squared given by the sum:
\begin{equation}\label{soam}
\chi_t^2=\chi_{SN}^{2}+\chi_{BAO}^{2}+\chi_{CMB}^{2}\,
\end{equation}

Our first fit leaves the coefficients free to vary, as reported in the first of Equations (\ref{paurzx}).~We therefore fixed the present Hubble rate to the upper value imposed by PLANCK   
 $\mathcal{H}_0=0.68\,h$, with \mbox{$h=100$ km s$^{-1}$ Mpc$^{-1}$}. Afterwards, we imposed the Hubble Space Telescope (HST) \mbox{measurement \cite{antesorla}} for $\mathcal H_0$. Finally, we performed a fit by considering the first order of the luminosity distance $d_L=\frac{1}{\mathcal{H}_0}z$ within the interval $z<0.36$ of SNeIa only, following the technique developed in \cite{chryorla}. Our cosmological priors are: $h\in[0.40,0.80], q_0\in[-1.50,1.00], j_0\in[-5.00,5.00]$ and $s_0\in[-20.00,20.00]$. The results are summarized in the {Table 1}.

Once the numerical bounds have been found, another possible advantage of cosmography is that one can get constraints of a certain model inverting the results of {Table 1}, obtaining the free coefficients of a given model in terms of the CS. In other words, let us consider that a given model leads to a vector of free parameters, the quantities $\xi_{i}$, with $i=1,\ldots, N$. Then, it would be possible to write down, by means of Equations (\ref{ciao}), the conditions:
\begin{eqnarray}\label{basis}
q_0=q_0\left(\xi_i\right) \,,\qquad
j_0=j_0\left(\xi_i\right) \,,\qquad
s_0=s_0\left(\xi_i\right) \,
\end{eqnarray}
and so, since the luminosity curves are thought to be one-to-one invertible, one gets:
\begin{eqnarray}\label{inbasis}
\xi_i=\xi_i\left(q_0\right) \,,\qquad
\xi_i=\xi_i\left(j_0\right) \,,\qquad
\xi_i=\xi_i\left(s_0\right) \,
\end{eqnarray}
Then, if $i\leq3$, it is possible to relate $\xi_i$ with the whole series; which is equivalent to having:
\begin{equation}\label{hyuuuuio}
\xi_i=\xi_i\left(q_0,j_0,s_0\right)\,
\end{equation}
For the sake of clarity, once the CS is known, to derive the corresponding errors on observables, one can invoke the use of logarithmic error propagation, discarding systematics, which has been already considered in the Union 2.1 compilation. In particular, assuming $\Psi\in\xi_i$, a generic parameter of a certain model, it follows $\Psi=\Psi(\alpha_{1},\ldots,\alpha_{N})$, with $\alpha_{i}$, $\forall i\in[1,\ldots,N]$. Hence, we have:
\begin{equation}\label{g}
\delta\Psi=\sum_{i}^{N}\frac{\partial \Psi}{\partial
\alpha_{i}}\times\delta\alpha_{i} \,
\end{equation}
giving:
$\Psi\pm\delta\Psi$
with $\delta\alpha_i$, the errors associated with the $i$-th variable.

It is worth noticing that a complete estimation of the total error associated with each parameter suffers from systematic uncertainties, which are notoriously hard to treat. However, recent PLANCK    
 surveys and the improved Union 2.1 compilation are able to significantly reduce systematics on measurements. It follows that overly optimistic estimates of systematic uncertainties do not modify the cosmographic analyses we have performed, and so, we do not account for a full description of systematical errors in \mbox{this review.}

The thus obtained experimental results appear to be actually dependent on particular priors imposed on the analyses. In particular, as is evident from {Table 1} and {Figure 1}, the first fit, in which no priors have been used, provides a more negative acceleration parameter than fits 2--4, showing an excellent agreement with theoretical predictions. A much smaller jerk parameter is also involved, and it is also present when $\mathcal{H}_0$ is fixed in the redshift interval $z<0.36$. This implies that the choice of a higher value for $\mathcal{H}_0$, than the one inferred from the PLANCK   
 mission, seems to be more comfortable in the framework of cosmography. This does not however contradict PLANCK     
 measurements, since the cosmological results showed in \cite{giorno} depend on the choice of the cosmological model, as the author claimed in the text. Moreover, the cosmological parameters have been also fixed at a $68\%$ confidence level, showing that future incoming PLANCK    
 results will be clearly able to better match the cosmographic intervals. The outcomes of fit 3 predict a slightly smaller acceleration parameter than fit 1 and a greater value for $j_0$. Overall, our results resemble much more the $\Lambda$CDM predictions, and agree with a higher value of the Hubble parameter than the one compatible with the $\Lambda$CDM model.

\begin{figure}[ht!]
\centering
\subfigure[]{
\includegraphics[width=2.35in]{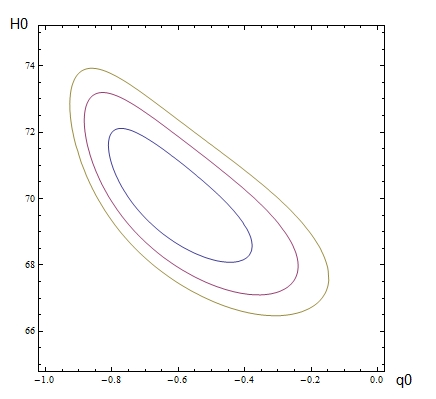}}
\subfigure[]{
\includegraphics[width=2.22in]{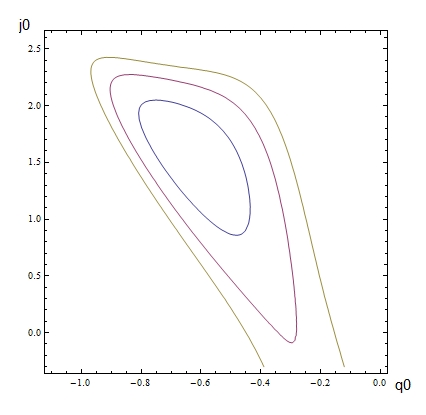}}
\subfigure[]{
\includegraphics[width=2.20in]{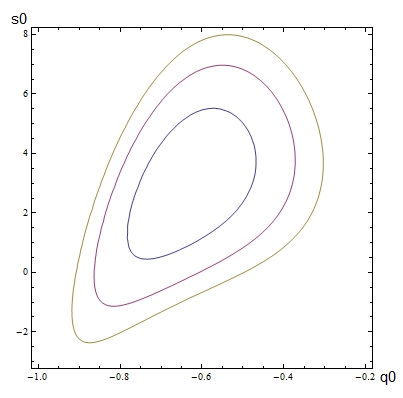}}
\subfigure[]{
\includegraphics[width=2.24in]{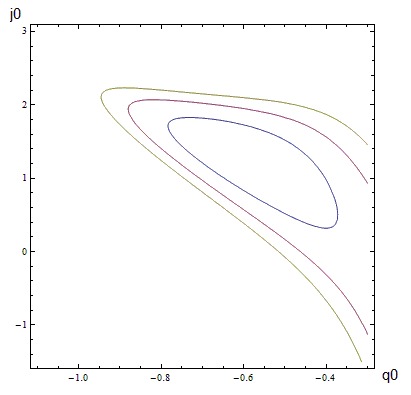}}
\subfigure[]{
\includegraphics[width=2.24in]{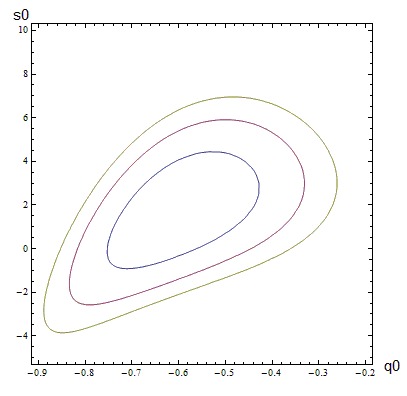}}
{\caption{Two-dimensional marginalized contour plots for the set A and B by using the combined tests with SNeIa, BAO and CMB. We used the luminosity distance as a function of the redshift, $z$, only. The last two figures below are based on set B with $h=0.6996$. We showed in the plots the $1\sigma$, $2\sigma$ and $3\sigma$ confidence levels, evaluated by means of standard Bayesian analyses on a grid.}
\label{Fig:EoS}}
\end{figure}

\newpage
The HST value seems to be disfavored, because of the higher $\mathcal{H}_0$, if compared to the cosmographic value, whereas the PLANCK    
 Hubble rate is also disfavored, providing a lower value, which is incompatible with theoretical predictions on CS. However, the acceleration parameter values are rather close to the theoretical values for fits 1--2--4, while both jerk and snap are significantly depending on the choice of $\mathcal{H}_0$, giving rise to non-conclusive results. The imposition of priors leads to rather disadvantageous outcomes, even for $\Omega_m$ evaluations. The predicted values of $j_0$ and $s_0$ do not constrain the Chaplygin gas fairly well, while fit 3 cannot constrain CPL. Further, the DGP and $\alpha$DGP models are not completely well bounded by cosmography, whereas the $\Lambda$CDM and $\omega$CDM are excellently constrained by the observations, resulting in them being compatible with recent limits \cite{solla.1, solla.2, solla.3}. Our results are also in close accordance with previous analyses and are able to exclude that the jerk parameter may be negative, showing that the concordance model predicts a transition at redshift $z\leq1$. In addition, we cannot also exclude \emph{a priori} that the jerk parameter is nearly identical to $j_0=1$, as predicted by the $\Lambda$CDM model, although it is apparently larger than the $\Lambda$CDM value. This leads to the consequence that $f(\mathcal R)$ and $f(\mathcal T)$ are also well constrained by cosmography, as one can see from Table 1. It turns out that a degeneracy problem occurs between modified theories of gravity and DE models, and cosmography is unable to remove these degeneracies using present data. More likely, the use of future surveys will permit us to discriminate between the modification of GR and DE models. To do so, a possibility is offered by better bounding the snap parameter, $s_0$. To get improved constraints on $s_0$, we need additional high redshift data. In fact, $s_0$ is the coefficient that enters the fourth orders of cosmographic Taylor expansions. Thus, future surveys with additional data will likely better constrain $s_0$, reducing the degeneracy problems between cosmological models.

Finally, it is worth noticing that by looking at Figure 1, our contours show that cosmographic constraints are fairly well bounded at the $68\%$ confidence level only. The additional $95\%$ and $98\%$ confidence levels provide error bars, which do not lead to accurate bounds on the CS. This is a consequence of the lack of current data. Again, future extended datasets would improve the quality of our fits and reduce error bars. This will permit us to determine precise limits on the cosmographic~coefficients.

\section{Conclusions and Perspectives} \label{sec conclusioni}

We reviewed the problem of DE through the use of the cosmography of the universe. In particular, we considered cosmography as a tool to select among competing models, showing that cosmography is, at most, a selection criterion to discriminate which model works fairly well over others.

In so doing, we reviewed several popular DE cosmological models, and we constrained them by cosmographic analyses, through the use of a combined test, involving SNeIa, BAO and CMB~measurements.

In particular, we investigated the main features of $\Lambda$CDM, $\omega$CDM, CPL parametrization, the DGP model and its extension, $\alpha$DGP, Chaplygin gas, $f(\mathcal R)$ and $f(\mathcal T)$ gravities. For each class of models, we reported the corresponding interpretation of cosmological parameters in terms of the CS, dealing with the current values of cosmographic parameters. With these considerations in mind, it is possible to recast any model parameter in term of the CS.

This is an essential feature to reach a cosmographic test of a particular class of models. In principle, the degeneracy problem among cosmological DE models can be alleviated if one fixes the cosmological observables in terms of the CS, while the degeneracy between extended theories of gravity and DE models is not healed by cosmography.

It is also important to say that cosmography is almost model independent once the scalar curvature is fixed or considered constant. In so doing, once fixing the CS, it is possible to reconstruct the dynamics of any cosmological model in terms of quantities that do not depend on some particular hypothesis.

To this end, standard cosmological datasets, SNeIa, BAO and CMB, have been adopted through the use of a Bayesian analysis and by minimizing the sum of each chi-squared term.
We finally showed that the Hubble fluid is currently accelerating within an acceleration parameter in the range $-0.60<q_0<-0.50$ and a jerk parameter $j_0>0$. Unfortunately, no definitive results can be inferred from the snap parameter, $s_0$, due to problems on convergence of truncated cosmographic series. The cosmographic results are also plagued by a severe dependence on the Hubble rate today, and so, more precise measurements of $\mathcal{H}_0$ today would improve the quality of cosmographic fits.

Future developments are aimed at improving the role of the cosmographic test to wider observational datasets. In particular, the PLANCK   
 measurements would better discriminate the confidence levels on the CS capable of fixing improved bounds on the observable universe in the upcoming future \cite{NewPlanck.1, NewPlanck.2, NewPlanck.3}.

\acknowledgements{Acknowledgments}

SC and MDL acknowledges the support of INFN Sez. di Napoli (Iniziative Specifiche TEONGRAV and QGSKY). OL is supported by the European PONa3 00038F1 KM3NET (INFN) Project.

\section*{\noindent Conflicts of Interest}
\vspace{12pt}

The authors declare no conflict of interest.





\bibliographystyle{mdpi}
\makeatletter
\renewcommand\@biblabel[1]{#1. }
\makeatother


\end{document}